\documentclass[
 amsmath,amssymb,
 aps,
 prb,
 superscriptaddress,
 twocolumn,
 ]{revtex4-2}

\usepackage{multirow}
\usepackage{bigdelim}
\usepackage{graphicx}
\usepackage{color}
\usepackage{dcolumn}
\usepackage{physics}
\usepackage{mathtools}
\usepackage{bm}
\usepackage{array}
\usepackage{comment}
\usepackage{cancel}
\usepackage{colortbl}
\usepackage{amsthm}

\newcommand{\floor}[1]{\left\lfloor#1\right\rfloor}
\newcommand{\ceil}[1]{\left\lceil#1\right\rceil}

\newcommand{\imi}{\mathrm{i}}

\newcommand{\poly}[3]{P_{#1}^{#2}(#3)}
\newcommand{\Dpoly}[3]{\partial P_{#1}^{#2}(#3)}

\newcommand{\tmat}[3]{\tau_{#2}(#3)}
\newcommand{\tmatN}[3]{\tau_{#1}^{#2}(#3)}

\newcommand{\tmattil}[3]{\tilde{\tau}_{#2}(#3)}
\newcommand{\Dtmat}[3]{\tau_{#2}'(#3)}

\newcommand{\modeset}[2]{\mathcal{S}^{#2}}
\newcommand{\modesetpositive}[2]{\mathcal{I}^{#2}}

\newcommand{\syk}[2]{H_{#2}}
\newcommand{\sykN}[2]{H_{#1}^{(#2)}}

\newcommand{\monodromyfwd}[3]{\overset{\rightarrow}{T}_{#2}(#3)}
\newcommand{\monodromybwd}[3]{\overset{\leftarrow}{T}_{#2}(#3)}

\newcommand{\monodromyfwdN}[3]{\overset{\rightarrow}{T}_{#1,#2}(#3)}

\newcommand{\isingH}[2]{H_{\mathrm{Ising}}^{#2}}

\def\linkcolor{BlueViolet}

\usepackage[usenames,dvipsnames]{xcolor}
\usepackage{hyperref}
\hypersetup{
    colorlinks=true, urlcolor=\linkcolor,citecolor=\linkcolor,linkcolor=\linkcolor
}

\newcommand{\tokyo}{\affiliation{Department of Physics, Graduate School of Science, The University of Tokyo, \\ 7-3-1, Hongo, Bunkyo-ku, Tokyo, 113-0033, Japan}}
\newcommand{\tokyoIntelligence}{\affiliation{Institute for Physics of Intelligence, The University of Tokyo, 7-3-1 Hongo, Tokyo 113-0033, Japan}}
\newcommand{\tokyoTrans}{\affiliation{Trans-Scale Quantum Science Institute, The University of Tokyo, 7-3-1, Hongo, Tokyo 113-0033, Japan}}

\preprint{APS/123-QED}

\begin{document}

\title{Integrability of a family of clean SYK models from the critical Ising chain}

\author{Kohei Fukai}
\email{kohei.fukai@phys.s.u-tokyo.ac.jp}
\tokyo
\author{Hosho Katsura}
\email{katsura@phys.s.u-tokyo.ac.jp}
\tokyo
\tokyoIntelligence
\tokyoTrans

\begin{abstract}
    \noindent
    We establish the integrability of a family of Sachdev-Ye-Kitaev (SYK) models with uniform $p$-body interactions.
    We derive the R-matrix and mutually commuting transfer matrices that generate the Hamiltonians of these models, and obtain their exact eigenspectra and eigenstates.
    Remarkably, the R-matrix is that of the critical transverse-field Ising chain.
    This work reveals an unexpected connection between the SYK model, central to many-body quantum chaos, and the critical Ising chain, a cornerstone of statistical mechanics.
\end{abstract}

\date{\today}

\maketitle

\section{Introduction}

The Sachdev-Ye-Kitaev (SYK) model~\cite{sy1993,kitaev,Maldacena-prx-2016,rosenhaus2019introduction,trunin,chowdhury2022sachdev,jha2025introduction} has emerged as a paradigmatic example of quantum many-body chaos, exhibiting maximal scrambling while remaining analytically tractable in the large-$N$ limit.
The model consists of $N$ Majorana fermions with random all-to-all $p$-body interactions and saturates the chaos bound~\cite{Maldacena2016jhep,kitaev,Polchinski2016, kobrin_syk_numerical}, making it a valuable theoretical laboratory for studying quantum chaos and its connection to black hole physics~\cite{sekino,Shenker2014}.
Moreover, the model exhibits non-Fermi liquid behavior in the large-$N$ limit~\cite{sachdevprx, chowdhury2022sachdev}, a feature shared by strange metals such as cuprates.

The essential role of disorder in the SYK model has motivated the search for simpler, disorder-free variants that retain the model's key features~\cite{Lau2021,ozaki-katsura-prr-2025, iyoda2018,witten2019syk,gurau2017complete,klebanov2017uncolored,krishnan2018contrasting,Wang2020,Balasubramanian2021,Claps,gorsky2025theta}.
In particular, Witten showed that tensor models can reproduce the same large-$N$ limit as the SYK model without quenched disorder~\cite{witten2019syk}.
While most disorder-free constructions aim to reproduce the chaotic features of the SYK model, it has been discovered that certain clean SYK models instead exhibit exact integrability.

Previous work demonstrated the integrability of specific clean SYK models.
The four-body clean SYK model with uniform couplings, first introduced in Ref.~\cite{Lau2021}, was shown to be exactly solvable~\cite{ozaki-katsura-prr-2025}, while the supersymmetric variant with a three-body supercharge was also solved explicitly~\cite{ozaki-katsura-prr-2025}.
Remarkably, despite exhibiting Poisson-like level spacing statistics characteristic of integrable systems, the out-of-time-order correlators (OTOCs) in these models show exponential growth at early times~\cite{ozaki-katsura-prr-2025}, a behavior typically associated with quantum chaos.
Additionally, some SYK variants with structured randomness, such as the Wishart SYK model, can be mapped to Richardson-Gaudin integrable models~\cite{iyoda2018}.
However, these integrable examples appeared as isolated cases, leaving open the question of whether there exists a unified framework for the integrability of clean SYK models.

In this work, we establish the integrability of SYK models with uniform (i.e., clean) $p$-body interactions.
While the SYK model is widely studied in the context of quantum chaos, the emphasis of the present work is on uncovering an integrable structure hidden in clean SYK models.
We construct an infinite family of mutually commuting SYK Hamiltonians, demonstrating that the previously studied models~\cite{Lau2021,ozaki-katsura-prr-2025} are special cases of this hierarchy.
Specifically, we show that the transfer matrix built from the R-matrix of the critical Ising chain encodes both nonlocal and local operators: when expanded in the spectral parameter, the coefficients of the transfer matrix yield all clean SYK Hamiltonians and the supercharges of their supersymmetric (SUSY) variants, whereas its logarithmic derivatives generate the Hamiltonian of the critical Ising chain and its local conserved charges.
The SYK Hamiltonians form a mutually commuting family, as do the supercharges among themselves, and each family commutes with the critical Ising Hamiltonian and its local conserved charges under appropriate boundary conditions.
This unexpected connection between the clean SYK models and the critical Ising chain provides a unified framework for understanding their exact solvability.

The key insight is that the integrability of these SYK models follows from the Yang-Baxter equation of the critical transverse-field Ising chain.
While the R-matrix for the critical Ising chain has been extensively studied in relation to the vertex model~\cite{davies1987yang, sierra-1993} or in the special representation of the Temperley-Lieb algebras~\cite{martin1991potts, temperley-lieb-rep-review, Miao-circuits, fukai-TL-2024}, a simple formulation in terms of Majorana fermions was only recently achieved~\cite{korepin-ising-yang-baxter-2025}.
This Majorana fermion representation for the R-matrix provides the crucial link to clean SYK models.

The paper is organized as follows.
In Sec.~\ref{sec:gen-syk}, we define the generalized clean SYK models and introduce their transfer matrices.
In Sec.~\ref{sec:integrability-from-ising}, we demonstrate how the integrability emerges from the critical Ising R-matrix, and we show that the SYK models and the critical Ising Hamiltonian belong to the same integrable family.
In Sec.~\ref{sec:solution-free-fermion-mode}, we obtain their exact eigenspectra and eigenstates.
We conclude in Sec.~\ref{sec:conclusion} with a discussion of our results and future perspectives.
The details of the proofs are provided in the Appendixes.

\section{Clean SYK models}\label{sec:gen-syk}
We first introduce a family of Hermitian operators in the clean SYK models with $p$-body interactions:
\begin{align}\label{eq:even-odd-SYK}
    \syk{N}{p} \equiv \imi^{\floor{p/2}} \sum_{1\leq i_1 < i_2 < \cdots < i_{p} \leq N} \gamma_{i_1} \gamma_{i_2} \cdots \gamma_{i_{p}}
    \,,
\end{align}
where $p \le N$ and $\gamma_i$ ($i=1, 2, \ldots, N$) are Majorana fermions satisfying $\{\gamma_i, \gamma_j\} = 2\delta_{i, j}$ and $\gamma_j^\dag = \gamma_j$.
We also define $\syk{N}{p} = 0$ for $p > N$.
The operators $\syk{N}{p}$ are Hermitian because of the factor $\imi^{\floor{p/2}}$.
In the following, we call $\syk{N}{p}$ the SYK charges.

We note that the odd-body SYK charges $\syk{N}{2p+1}$ can be seen as
supercharges of the clean counterpart of the disordered $\mathcal{N}=1$ SUSY SYK model introduced in~\cite{sachdevprd2017susy}.
We also note that $\syk{N}{3}$ corresponds to the supercharge studied in~\cite{ozaki-katsura-prr-2025}, while $\syk{N}{4}$ is the clean SYK Hamiltonian discussed in~\cite{Lau2021, ozaki-katsura-prr-2025}.
Hereafter, we refer to $\syk{N}{2p}$ as SYK Hamiltonians and $\syk{N}{2p+1}$ as SYK supercharges.

We define a one-parameter family of transfer matrices for the SYK Hamiltonians as
\begin{align}
    \tmat{N}{+}{u}
    =
    \sum_{p=0}^{\floor{N/2}} (-u)^{p} \syk{N}{2p}
    \,,
\end{align}
where $H_{0} \equiv I$ is the identity operator.
These transfer matrices 
commute for different values of the parameters:
\begin{align}
    \label{eq:transfer-mat-mutual-commutativity}
    [\tmat{N}{+}{u}, \tmat{N}{+}{v}] = 0
    \,,
\end{align}
which immediately leads to the mutual commutativity of the SYK Hamiltonians:
\begin{align}
    \label{eq:even-commute}
    [\syk{N}{2p}, \syk{N}{2p'}]     & = 0\,,
\end{align}
where $p$ and $p'$ are nonnegative integers.

For the SYK supercharges, we define the transfer matrix as
\begin{align}
    \tmat{N}{-}{u}
    =
    \sum_{p=0}^{\floor{(N-1)/2}} (-u)^{p} \syk{N}{2p+1}
    \,.
\end{align}
These transfer matrices are mutually commuting:
\begin{align}
    \label{eq:transfer-mat-mutual-commutativity-odd}
    [\tmat{N}{-}{u}, \tmat{N}{-}{v}] = 0
    \,,
\end{align}
which immediately leads to the mutual commutativity of the SYK supercharges:
\begin{align}
    \label{eq:odd-commute}
    [\syk{N}{2p+1}, \syk{N}{2p'+1}] & = 0\,,
\end{align}
where $p$ and $p'$ are nonnegative integers.
The proof of Eqs.~\eqref{eq:transfer-mat-mutual-commutativity} and~\eqref{eq:transfer-mat-mutual-commutativity-odd} is explained from the integrability of the critical Ising chain in the next section.

The SYK Hamiltonians and supercharges are related through the anticommutator with the first supercharge $\syk{N}{1} = \sum_{j=1}^{N} \gamma_j$:
\begin{align}\label{eq:even-odd-relation}
    \frac{1}{2} \{\syk{N}{2p}, \syk{N}{1}\} = \syk{N}{2p+1}
    \,,
\end{align}
which can be proved by Eq.~\eqref{eq:H-gamma-comm2} in Appendix~\ref{app:tau-fermion-relation}.

\section{Integrability from critical Ising chain}\label{sec:integrability-from-ising}
Here, we will show that the clean SYK models are integrable, which follows from the integrability of the critical Ising chain.

The R-matrix for the critical Ising chain~\cite{korepin-ising-yang-baxter-2025} is given by
\begin{align}
    \label{eq:R-mat}
    R_{a,j}(u) = \gamma_a - u \gamma_j
    \,,
\end{align}
where $u$ is the spectral parameter, and $\gamma_a$ is the auxiliary Majorana fermion satisfying $\{\gamma_a, \gamma_j\} = 0 \ (1\le j \le N)$, $\gamma_a^2 = 1$ and $\gamma_a^\dag = \gamma_a$.
The R-matrix satisfies the Yang-Baxter equation:
\begin{align}
    \label{eq:YBE}
    R_{a,b}(u / v) R_{a,j}(u) R_{b,j}(v) = R_{b,j}(v) R_{a,j}(u)  R_{a,b}(u / v)
    \,,
\end{align}
where we also introduce the second auxiliary Majorana fermion $\gamma_b$.
The inversion relation for the R-matrix is
\begin{align}\label{eq:R-mat-inv}
    R_{a,j}(u)^2 = 1 + u^2
    \,.
\end{align}
Unlike the conventional R-matrix with difference form $R(u,v) = R(u-v)$, our R-matrix~\eqref{eq:R-mat} takes the multiplicative form $R(u,v) = R(u/v)$.

We note that Eq.~\eqref{eq:YBE} is the nonbraided Yang-Baxter equation with a nonlocal R-matrix in terms of Majorana fermions.
This differs from the braided formulation of an R-matrix studied extensively in the literature~\cite{Ivanov2001, Kauffman2004braiding, Kauffman-2016, Kauffman-2018}.
The nonbraided and nonlocal formulation of the R-matrix using Majorana fermions is the new perspective in Ref.~\cite{korepin-ising-yang-baxter-2025}.

We define the forward and backward monodromy matrices:
\begin{align}
    \label{eq:monodromy-fwd}
    \monodromyfwd{N}{a}{u} & \equiv \prod_{j=1}^{N} R_{a,j}((-1)^{j} \sqrt{\imi u})
    \,,                                                                             \\
    \label{eq:monodromy-bwd}
    \monodromybwd{N}{a}{u} & \equiv \prod_{j=N}^{1} R_{a,j}((-1)^{j} \sqrt{\imi u})
    \,.
\end{align}
Here, the forward product~\eqref{eq:monodromy-fwd} is ordered from $j=1$ to $N$ (left to right), while the backward product~\eqref{eq:monodromy-bwd} is from $j=N$ to $1$.
These monodromy matrices satisfy the RTT relation:
\begin{align}
    \label{eq:RTT}
    R_{a,b}(\sqrt{u / v}) \monodromyfwd{N}{a}{u} \monodromyfwd{N}{b}{v} = \monodromyfwd{N}{b}{v} \monodromyfwd{N}{a}{u}  R_{a,b}(\sqrt{u / v})
    \,,
\end{align}
and the same relation holds for the backward monodromy matrix $\monodromybwd{N}{a}{u}$.
Equation~\eqref{eq:RTT} can be proved using the Yang-Baxter equation~\eqref{eq:YBE} repeatedly:
\begin{widetext}
    \begin{align}
          & R_{a,b}(\sqrt{u / v}) \monodromyfwd{N}{a}{u} \monodromyfwd{N}{b}{v} = (-1)^{N(N-1)/2} R_{a,b}(\sqrt{ \imi u} / \sqrt{\imi v}) \prod_{j=1}^{N} R_{a,j}((-1)^{j} \sqrt{\imi u}) R_{b,j}((-1)^{j} \sqrt{\imi v})
        \nonumber                                                                                                                                                                                                         \\
        = & (-1)^{N(N-1)/2} R_{b,1}(-\sqrt{\imi v}) R_{a,1}(- \sqrt{\imi u}) R_{a,b}(\sqrt{ \imi u} / \sqrt{\imi v}) \prod_{j=2}^{N} R_{a,j}((-1)^{j} \sqrt{\imi u}) R_{b,j}((-1)^{j} \sqrt{\imi v})
        \nonumber                                                                                                                                                                                                         \\
        = & \cdots =
        (-1)^{N(N-1)/2} \qty[ \prod_{j=1}^{N} R_{b,j}((-1)^{j} \sqrt{\imi v}) R_{a,j}((-1)^{j} \sqrt{\imi u})] R_{a,b}(\sqrt{ \imi u} / \sqrt{\imi v})
        \nonumber                                                                                                                                                                                                         \\
        = & \monodromyfwd{N}{b}{v} \monodromyfwd{N}{a}{u}  R_{a,b}(\sqrt{u / v})
        \,.
        \nonumber
    \end{align}
\end{widetext}

The key observation is that these monodromy matrices decompose into the SYK transfer matrices:
\begin{align}
    \label{eq:monodromy-matrix-fwd}
    \monodromyfwd{N}{a}{u} & =\qty( \tmat{N}{+}{u} - \gamma_a \sqrt{\imi u} \tmat{N}{-}{u}) \gamma_a^{N}
    \,,                                                                                                       \\
    \label{eq:monodromy-matrix-bwd}
    \monodromybwd{N}{a}{u} & = \gamma_a^{N} \qty(  \tmat{N}{+}{-u} + \gamma_a \sqrt{\imi u} \tmat{N}{-}{-u} )
    \,.
\end{align}
In Appendix~\ref{app:monodromy-matrix-fwd}, we prove Eq.~\eqref{eq:monodromy-matrix-fwd} by induction using the recursion for the transfer matrix~\eqref{eq:transfer-mat-recursion}.
Equation~\eqref{eq:monodromy-matrix-bwd} can also be proved similarly.

Substituting Eq.~\eqref{eq:monodromy-matrix-fwd} into the RTT relation~\eqref{eq:RTT}, we can prove the mutual commutativity of the transfer matrices~\eqref{eq:transfer-mat-mutual-commutativity} and~\eqref{eq:transfer-mat-mutual-commutativity-odd}, thereby establishing the mutual commutativity of the SYK charges in Eqs.~\eqref{eq:even-commute} and \eqref{eq:odd-commute}.
The detailed proof is given in Appendix~\ref{app:transfer-mat-mutual-commutativity}.

From the inversion relation~\eqref{eq:R-mat-inv}, we can see that the product of forward and backward monodromy matrices becomes
\begin{align}
    \label{eq:monodromy-inversion}
    \monodromyfwd{N}{a}{u} \monodromybwd{N}{a}{u} = \monodromybwd{N}{a}{u} \monodromyfwd{N}{a}{u} = (1 + \imi u)^N
    \,.
\end{align}
Substituting Eqs.~\eqref{eq:monodromy-matrix-fwd} and \eqref{eq:monodromy-matrix-bwd} into Eq.~\eqref{eq:monodromy-inversion}, we obtain
\begin{align}
    \label{eq:char-poly}
    \poly{N}{\pm}{u^2} \equiv \tmat{N}{\pm}{u} \tmat{N}{\pm}{-u} = \frac{(1 + \imi u)^N \pm (1 - \imi u)^N}{(1 + \imi u) \pm (1 - \imi u)}\,.
\end{align}
Using the substitution $u=\tan(\kappa/2)$, the polynomials become
\begin{align}
     \label{eq:char-poly-tan-plus}
    \poly{N}{+}{u^2} &= \frac{\cos(N \kappa / 2)}{\cos^{N}(\kappa/2)}\,,
    \\
      \label{eq:char-poly-tan-minus}
    \poly{N}{-}{u^2} &= \frac{\sin(N \kappa / 2)}{\cos^{N-1}(\kappa/2) \sin(\kappa/2)}\,.
\end{align}
The proof of Eq.~\eqref{eq:char-poly} is given in Appendix~\ref{app:char-poly}.
In Sec.~\ref{sec:solution-free-fermion-mode}, we show that Eq.~\eqref{eq:char-poly} is the characteristic polynomial determining the spectra of the SYK charges.

From the factorized form of the monodromy matrices~\eqref{eq:monodromy-fwd} and~\eqref{eq:monodromy-bwd}, we can easily calculate the conjugation of a single Majorana fermion with the monodromy matrices, and then with the transfer matrices.
Here we give the final result, and the detailed derivation is given in Appendix~\ref{app:gamma1-conjugation}:
\begin{align}
     &\tmat{N}{+}{u} \gamma_1 \tmat{N}{+}{-u}
     =
     \frac{1}{\cos^{N}(\kappa/2)}
      \nonumber\\
     &\quad
     \times
     \biggl[
    \cos\biggl(\biggl(\frac{N}{2}-1\biggr)\kappa\biggr)
    \gamma_1
    +
    \imi \sin\kappa
    \sum_{l = 2}^{N} e^{-\imi (N/2+1-l)\kappa} \gamma_{l}
    \biggr]
    \,,
    \label{eq:gamma1-conjugation-even}
    \\
     &\tmat{N}{-}{u} \gamma_1 \tmat{N}{-}{-u}
     =
     \frac{1}{\sin(\kappa/2) \cos^{N-1}(\kappa/2)}
     \nonumber\\
     &\quad
     \times
     \biggl[
    - \sin\biggl(\biggl(\frac{N}{2}-1\biggr)\kappa\biggr)
    \gamma_1
    +
    \sin\kappa
    \sum_{l = 2}^{N} e^{-\imi (N/2+1-l)\kappa} \gamma_{l}
    \biggr]
    \,,
    \label{eq:gamma1-conjugation-odd}
\end{align}
where again $u = \tan(\kappa/2)$.
Using the translation operators $\tmat{N}{\pm}{-\imi}$, which will be explained below in Eqs.~\eqref{eq:twisted-translation-even} and~\eqref{eq:twisted-translation-odd}, we can also obtain the other cases of the conjugation: $\tmat{N}{\pm}{u} \gamma_j \tmat{N}{\pm}{-u}$.

Having established the mutual commutativity of the SYK charges through the RTT relation, we now reveal an unexpected connection: the same transfer matrix also generates the critical Ising Hamiltonian.
This connection arises from the logarithmic derivative of the transfer matrix, which yields local conserved charges.
As a consequence, the SYK charges commute with the critical Ising Hamiltonian and its higher-order local conserved charges.
The critical Ising Hamiltonian in terms of Majorana fermions is given by
\begin{align}
    \label{eq:ising-in-Majorana}
    \isingH{N}{\pm} =  \imi \sum_{j=1}^{N-1} \gamma_j \gamma_{j+1} \pm \imi \gamma_N \gamma_1
    \,.
\end{align}
The choice of the sign ($\pm$) corresponds to periodic ($+$) or anti-periodic ($-$) boundary conditions on the Majorana fermions. 
The Hamiltonian $\isingH{N}{+}$ commutes with the supercharges $\syk{N}{2p+1}$, whereas $\isingH{N}{-}$ commutes with the SYK Hamiltonians $\syk{N}{2p}$, which will be proved below.
Via the Jordan-Wigner transformation $\gamma_{2j-1} = \qty(\prod_{l=1}^{j-1} Z_{l}) X_{j}$ and $\gamma_{2j} = \qty(\prod_{l=1}^{j-1} Z_{l}) Y_{j}$, where $X_j$, $Y_j$, and $Z_j$ are the Pauli matrices acting on the $j$th site, we have
\begin{align}
    \label{eq:ising-in-spin}
    \isingH{N}{\pm} = -\sum_{j=1}^{\ceil{N/2}-1} X_j X_{j+1} - \sum_{j=1}^{\floor{N/2}} Z_j \pm P h_{\mathrm{bdry}}\,,
\end{align}
where $P \equiv Z_1 Z_2 \cdots Z_{\ceil{N/2}}$ and the boundary term $h_{\mathrm{bdry}}$ is given by
\begin{align}
    h_{\mathrm{bdry}}
    \equiv
    \begin{cases}
        X_{N/2} X_{1} & (\text{for even $N$})
        \\
        - Y_{(N+1)/2} X_{1} & (\text{for odd $N$})
    \end{cases}
    \,.
\end{align}
Here, we have constructed the representation of Majorana fermions as operators on the Hilbert space $(\mathbb{C}^2)^{\otimes \ceil{N/2}}$.

The $\pm P$ term in Eq.~\eqref{eq:ising-in-spin} has a deep connection to noninvertible symmetries of the critical Ising chain~\cite{korepin-ising-yang-baxter-2025, Seiberg-Shao2024}.
The operators $(1 \pm P)$ act as projectors onto different parity sectors, and any conserved quantity $Q$ of our system can be used to construct noninvertible symmetries $Q_\pm = Q(1 \pm P)$.

In particular, the Kramers-Wannier duality operator, which becomes a symmetry at criticality~\cite{korepin-ising-yang-baxter-2025,zhu2025}, takes the form $\mathsf{D} = \mathcal{U}(1+P)$, where $\mathcal{U}$ is the twisted translation operator~\cite{seiberg2025lsmcpt} corresponding to the transfer matrix at $u=-\imi$:
\begin{align}
    \label{eq:twisted-translation-op}
    \mathcal{U}
    =
    \tmat{N}{+}{-\imi}
    =
    \qty(\prod_{j=2}^{N-1}
    P_{1,j}^{(-)^j}) \gamma_1^{N-1}
    \,.
\end{align}
Here we defined $P_{j,l}^{\pm} \equiv \gamma_j \pm \gamma_l$.
The twisted translation acts as
\begin{align}
    \label{eq:twisted-translation-even}
    \mathcal{U} \gamma_j \mathcal{U}^{-1}
    =
    \begin{cases}
        - \gamma_{j-1} & (j > 1)\\
        \gamma_{N} & (j = 1)
    \end{cases}\,.
\end{align}
The translation generated by the transfer matrix for the supercharges is just a simple translation~\cite{Hsieh2016,Sannomiya-susy-2019,piroli2021fermionic}:
\begin{align}
    \label{eq:twisted-translation-odd}
    \mathcal{U}' \gamma_j (\mathcal{U}')^{-1}
    =
    \gamma_{j-1}\,,
\end{align}
where $\mathcal{U}' \equiv \tmat{N}{-}{-\imi} = \tmat{N}{+}{-\imi} \gamma_1$ and the indices are taken modulo $N$.
Equations~\eqref{eq:twisted-translation-op}-\eqref{eq:twisted-translation-odd} can be proved using a similar argument to that in Ref.~\cite{korepin-ising-yang-baxter-2025}.
We note that the action of the twisted translation~\eqref{eq:twisted-translation-even} differs from that in Ref.~\cite{korepin-ising-yang-baxter-2025}.
This is because our monodromy matrix is constructed from the staggered choice of spectral parameters in Eqs.~\eqref{eq:monodromy-fwd} and \eqref{eq:monodromy-bwd}, whereas the monodromy matrix in Ref.~\cite{korepin-ising-yang-baxter-2025} uses a uniform choice of spectral parameters.

The critical Ising Hamiltonian~\eqref{eq:ising-in-Majorana} can be derived from the logarithmic derivative of the transfer matrix at $u=-\imi$:
\begin{align}
    \label{eq:log-derivative-ising}
    \pdv{u}\eval{\ln \tmat{N}{\pm}{u}}_{u=-\imi}
     & =
    - \frac{1}{4} \isingH{N}{\mp}  + C_{\pm}
    \,,
\end{align}
where $C_{\pm} = \imi (N \pm 1 - 1)/4$.
The proof of Eq.~\eqref{eq:log-derivative-ising} is given in Appendix~\ref{app:log-derivative-ising}.
Equation~\eqref{eq:log-derivative-ising} means that the SYK charges with even/odd Majorana fermions commute with the critical Ising Hamiltonian $\isingH{N}{\mp}$, and then can be simultaneously diagonalized.
Higher-order derivatives of the logarithm of the transfer matrix give the higher-order local conserved charges in the Ising chain~\cite{Dolan-charge-1982,Grady-ising-charge-1982,Prosen-1998,Fagotti-2013}, which are all bilinear in Majorana fermions, as can be easily seen from Eqs.~\eqref{eq:tau-fermion-prod} and~\eqref{eq:tau-fermion-prod-odd} below.

\section{Exact solution}\label{sec:solution-free-fermion-mode}
Here, we give the exact eigenspectra and eigenstates 
of the clean SYK charges.
We first define the 
fermionic annihilation operators $f_k$ as the Fourier transforms of the Majorana fermions $\gamma_j$~\cite{ozaki-katsura-prr-2025}:
\begin{align}\label{eq:fermion-mode}
    f_{k} = \frac{1}{\sqrt{2N}}\sum_{j=1}^{N} e^{\imi   (j-1)k } \gamma_j
    \quad
    (k \in \modesetpositive{N}{\pm})
    \,,
\end{align}
where $\modesetpositive{N}{+} \equiv \left\{ k \in \frac{2 \pi}{N} \qty(\mathbb{Z} + \frac{1}{2}) : 0 < k < \pi\right\}$ and $\modesetpositive{N}{-} \equiv \left\{ k \in \frac{2 \pi}{N} \mathbb{Z} : 0 < k < \pi\right\}$.
The fermionic creation operators are then defined by $f_k^\dag$. 
These complex fermion operators satisfy the anti-commutation relations: $\{f_k^\dag, f_{k'} \} = \delta_{k,k'}$, $\{ f_k, f_{k'} \} = \{f^\dagger_k, f^\dagger_{k'}\}=0$.
We also define the Majorana zero mode: $\chi_0 \equiv \frac{1}{\sqrt{N}}\sum_{j=1}^N \gamma_j$ and the Majorana $\pi$ mode: $\chi_\pi \equiv \frac{1}{\sqrt{N}}\sum_{j=1}^N (-1)^{j-1} \gamma_j$, satisfying $\chi_0^2 = \chi_\pi^2 = 1$ and $\{\chi_0, \chi_\pi\} = 0$. 
Also, they satisfy $\{\chi_0, f_k\} = 0$ for $k \in \modesetpositive{N}{-}$ and $\{\chi_\pi, f_k\} = 0$ for $k \in \modesetpositive{N}{(-)^{N+1}}$.

Using the complex fermions above, the transfer matrices are diagonalized as
\begin{align}
    \label{eq:tau-fermion-prod}
    \tmat{N}{+}{u} & = \prod_{k \in \modesetpositive{N}{+}}
    \left\{1 - u \epsilon_k \left( n_k - \frac{1}{2}  \right) \right\}
    \,,
    \\
    \label{eq:tau-fermion-prod-odd}
    \tmat{N}{-}{u} & = \sqrt{N}\chi_0 \prod_{k \in \modesetpositive{N}{-}}
    \left\{1 - u \epsilon_k \left( n_k - \frac{1}{2}  \right) \right\}
    \,,
\end{align}
where $n_k \equiv f_k^\dag f_{k}$ is the number operator for the mode of $k$ and $\epsilon_k$ is given by
\begin{align}
    \label{eq:single-particle-energy}
    \epsilon_k & = 2\cot\left(\frac{k }{2}\right)
    \,.
\end{align}
Note that $\epsilon_k$ are the single-particle energies of the quadratic Hamiltonians: $\syk{N}{2}$ for $k \in \modesetpositive{N}{+}$ and $\sqrt{N}\chi_0 \syk{N}{3}$ for $k \in \modesetpositive{N}{-}$~\cite{ozaki-katsura-prr-2025}.
These quadratic forms enabled the diagonalization of $\syk{N}{4}$ and $(\syk{N}{3})^2$ with the complex fermions $f_k$ in~\cite{ozaki-katsura-prr-2025}; here we extend this to all SYK charges $\syk{N}{p}$.
The eigenvalues of $\tmat{N}{\pm}{u}$ are thus given by $(\sqrt{N} s_0)^{\delta_\pm} \prod_{k \in \modesetpositive{N}{\pm}}(1 - u \cot(k /2) s_k)$ where $s_k \in \{+1, -1\}$, $\delta_{+} \equiv 0$ and $\delta_{-} \equiv 1$.
The proof of Eqs.~\eqref{eq:tau-fermion-prod} and~\eqref{eq:tau-fermion-prod-odd} is given in Appendix~\ref{app:tau-fermion-prod}.
The commutation relations of $\chi_\pi$ with the transfer matrices are evident from Eqs.~\eqref{eq:tau-fermion-prod} and~\eqref{eq:tau-fermion-prod-odd}: $\chi_\pi$ commutes with $\tmat{N}{+}{u}$ for odd $N$ but anticommutes with $\tmat{N}{-}{u}$ for even $N$.

The single-particle energies~\eqref{eq:single-particle-energy} are determined by the zeros of the polynomial Eq.~\eqref{eq:char-poly}:
\begin{align}
    \poly{N}{\pm}{u_k^2} = 0 \quad  (k \in \modesetpositive{N}{\pm})\,,
\end{align}
where $u_k \equiv 2/\epsilon_k = \tan(k/2)$.
This follows immediately from Eqs.~\eqref{eq:char-poly-tan-plus} and \eqref{eq:char-poly-tan-minus}.
Then, the polynomials are expressed as
\begin{align}
    \poly{N}{\pm}{u^2} & = \prod_{k \in \modesetpositive{N}{\pm}} (1 - u^2 / u_k^2)
    \,.
\end{align}

The eigenstates of the transfer matrices~\eqref{eq:tau-fermion-prod} can also be constructed.
Here, we only consider the even $N$ case.
The odd $N$ case is notoriously confusing~\cite{Seiberg-Shao2024,seiberg2025lsmcpt} and is not considered here.
For $\tmat{N}{+}{u}$, we denote by $\ket{0}$ the vacuum for the annihilation operators $f_k$, which satisfies $f_k \ket{0} = 0$ for all $k \in \modesetpositive{N}{+}$. 
Then, all eigenstates can be obtained by applying creation operators to the vacuum: $\prod_{q=1}^{n}f_{k_q}^\dagger \ket{0}$ for $k_q \in \modesetpositive{N}{+}$ and $n \le N/2$.

For $\tmat{N}{-}{u}$, its eigenstates are constructed from twofold degenerate vacua $\ket{0}_\pm$, which are also eigenstates of the Majorana zero mode: $\chi_0 \ket{0}_\pm = \pm \ket{0}_\pm$.
We note that here the eigenstates of $\tmat{N}{-}{u}$ do not simultaneously diagonalize the fermion parity operator $(-1)^F$.
Then, all eigenstates can be obtained by applying creation operators to the degenerate vacua: $\prod_{q=1}^{n}f_{k_q}^\dagger \ket{0}_\pm$ for $k_q \in \modesetpositive{N}{-}$ and $n < N/2$.
For the SUSY Hamiltonian made from the square of the supercharge, the eigenstates can be constructed to simultaneously diagonalize the fermion parity operator \cite{Witten1982}.
We discuss this point in Appendix~\ref{app:eigenstate-susy-hamiltonian}.
Also, please refer to Ref.~\cite{ozaki-katsura-prr-2025} for details.

By expanding Eqs.~\eqref{eq:tau-fermion-prod} and~\eqref{eq:tau-fermion-prod-odd}, we have the spectral decompositions of the SYK charges as
\begin{align}
    \syk{N}{2p} & = \sum_{\substack{\mathcal{K} \subseteq \modesetpositive{N}{+} \\ |\mathcal{K}| = p}} \prod_{k \in \mathcal{K}} \epsilon_{k} \left( n_{k} - \frac{1}{2} \right)
    \,,
    \label{eq:hamiltonian-supercharge-in-fermion}
    \\
    \syk{N}{2p+1} & = \sqrt{N} \chi_0 \sum_{\substack{\mathcal{K} \subseteq \modesetpositive{N}{-} \\ |\mathcal{K}| = p}} \prod_{k \in \mathcal{K}} \epsilon_{k} \left( n_{k} - \frac{1}{2} \right)
    \,,
    \label{eq:hamiltonian-supercharge-in-fermion-odd}
\end{align}
where $\mathcal{K}$ is a subset of $\modesetpositive{N}{\pm}$ with cardinality $p$.
The 
spectrum of $\syk{N}{p}$ is given by $(\sqrt{N} s_0)^{\delta_{\pm}}\sum_{\mathcal{K} \subseteq \modesetpositive{N}{(-)^p},\ |\mathcal{K}| = \floor{p/2}} \prod_{k \in \mathcal{K}} \cot(k / 2) s_{k}$ where $s_k \in \{+1, -1\}$, $\delta_{+} \equiv 0$ and $\delta_{-} \equiv 1$.
For odd $p$, the presence of the Majorana zero mode $\chi_0$ leads to the freedom to choose $s_0 = \pm 1$.
We note in passing that the form of the Hamiltonian $\syk{N}{2p}$ in Eq.~\eqref{eq:hamiltonian-supercharge-in-fermion} resembles that of the commuting SYK model discussed in \cite{gao2024commuting}.

We give some important ingredients for the proof of the exact solutions~\eqref{eq:tau-fermion-prod} and~\eqref{eq:tau-fermion-prod-odd}.
We first report the relationship between the complex fermion~\eqref{eq:fermion-mode} and the transfer matrix as
\begin{align}\label{eq:fermion-FFD-like}
    f_k
     & =
    \frac{e^{\imi  (j-1) k}}{\mathcal{N}_{k}}\tmat{N}{\pm}{u_k} \gamma_j \tmat{N}{\pm}{-u_k}
    \quad
    (k \in \modesetpositive{N}{\pm})
    \,,
\end{align}
where $\mathcal{N}_{k}$ is a normalization factor given in Eq.~\eqref{eq:normalization-factor} and $j \in \{1,\ldots,N\}$.
Equation~\eqref{eq:fermion-FFD-like} can be proved using Eqs.~\eqref{eq:gamma1-conjugation-even} and~\eqref{eq:gamma1-conjugation-odd}, whose details are given in Appendix~\ref{app:fermion-FFD-like}, and leads to the important identity for the proof of the exact solutions~\eqref{eq:tau-fermion-prod} and~\eqref{eq:tau-fermion-prod-odd}:
\begin{align}
    \label{eq:tau-fermion-relation}
    (u_k - u) \tmat{N}{\pm}{u} f_k & = \pm (u_k + u) f_k \tmat{N}{\pm}{u}
    \quad
    (k \in \modesetpositive{N}{\pm})
    \,.
\end{align}
The details of the proof of Eq.~\eqref{eq:tau-fermion-relation} are given in Appendix~\ref{app:tau-fermion-relation}.

We briefly sketch how we can prove Eq.~\eqref{eq:tau-fermion-relation}.
The conjugation of the complex fermion with the transfer matrices can be calculated as
\begin{align}
    \tmat{N}{\pm}{u} f_k \tmat{N}{\pm}{-u}
    & =
    \frac{1}{\mathcal{N}_{k}}
    \tmat{N}{\pm}{u} \tmat{N}{\pm}{-u_k} \gamma_1 \tmat{N}{\pm}{u_k} \tmat{N}{\pm}{-u}
    \nonumber\\
    &=
    \frac{1}{\mathcal{N}_{k}}
    \tmat{N}{\pm}{-u_k} ( \tmat{N}{\pm}{u} \gamma_1  \tmat{N}{\pm}{-u} ) \tmat{N}{\pm}{u_k}
    \nonumber\\
    &=
    \frac{1}{\mathcal{N}_{k}}
    \sum_{j = 1}^{N}
    c_j(u)
    \tmat{N}{\pm}{-u_k} \gamma_j  \tmat{N}{\pm}{u_k}
    \nonumber\\
    &=
    \qty(
    \sum_{j = 1}^{N}
    c_j(u) e^{\imi  (j-1) k}
    )
    f_k
    \,,
    \nonumber
\end{align}
where in the first equality, we used Eq.~\eqref{eq:fermion-FFD-like}; in the second equality, we used the mutual commutativity of the transfer matrices~\eqref{eq:transfer-mat-mutual-commutativity} and~\eqref{eq:transfer-mat-mutual-commutativity-odd}; in the third equality, we used Eqs.~\eqref{eq:gamma1-conjugation-even} and~\eqref{eq:gamma1-conjugation-odd}, and the coefficient $c_j(u)$ is read from Eqs.~\eqref{eq:gamma1-conjugation-even} and~\eqref{eq:gamma1-conjugation-odd}; and in the last equality, we used Eq.~\eqref{eq:fermion-FFD-like} again.
Multiplying both sides by $\tmat{N}{\pm}{u}$ from the right, we obtain Eq.~\eqref{eq:tau-fermion-relation}.

Equation~\eqref {eq:fermion-FFD-like} is similar to the construction of the fermion operators in the free fermions in disguise~\cite{disguise-Fendley-2019,FFD-chapman-2021,FFD-chapman-unified-2023, sajat-FP-model, Istvan-ffd-prb-2025, sajat-floquet, fukai-claw-2025, Eric-FFD-hilbert-space-2025, David-ffd-circ-2025}.
The $\gamma_j$ here play the same role as edge operators.
Here we have a rigorous relationship between the complex fermion constructed from the diagonalization of the single-particle Hamiltonian and that constructed from the way of free fermions in disguise.
Equation~\eqref{eq:fermion-FFD-like} is also important in the proof of Eqs.~\eqref{eq:tau-fermion-prod} and~\eqref{eq:tau-fermion-prod-odd}.

From Eqs.~\eqref{eq:log-derivative-ising} and~\eqref{eq:tau-fermion-prod}, the critical Ising Hamiltonian is also expressed in terms of the complex fermions as
\begin{align}
    \label{eq:ising-in-fermion-mode}
    \isingH{N}{\pm}
    = 4\sum_{k \in \modesetpositive{N}{\pm}} \sin k \left( n_k - \frac{1}{2} \right)
    \,.
\end{align}
The higher-order local conserved charges of the critical Ising chain have similar expressions.

\section{Conclusion}\label{sec:conclusion}
We have established the complete integrability of the clean SYK models with $p$-body interactions by revealing their connection to the critical transverse field Ising chain.
The key discovery is that the generalized SYK Hamiltonians emerge from transfer matrices constructed using the R-matrix of the critical Ising chain, which satisfies the Yang-Baxter equation.
Our approach provides explicit solutions for all clean SYK models, extending the previously known results~\cite{ozaki-katsura-prr-2025} for $\syk{N}{4}$ and $(\syk{N}{3})^2$ to the general $p$-body case $\syk{N}{p}$.
This unexpected connection between the family of clean SYK models and the critical Ising chain places the former within the well-established framework of Yang-Baxter integrability.

The framework developed here may also be extended to compute correlation functions and OTOCs in clean SYK models.
While these quantities were computed for $\syk{N}{4}$ and $(\syk{N}{3})^2$ in Ref.~\cite{ozaki-katsura-prr-2025}, the extension to general $p$-body interactions remains an open problem.

Furthermore, our exact solution naturally extends to hybrid models that combine SYK charges with local conserved charges of the critical Ising chain.
Such models, with Hamiltonians that are linear combinations of long-range and short-range interaction terms, have also been investigated in the literature~\cite{chen2020replica}, where the SYK model and the Kitaev chain are coupled.
Our results may provide new insights into the nature of such models.

\begin{acknowledgments}
    The authors thank Bal\'azs Pozsgay, Tam\'as Gombor, and Yuan Miao for fruitful discussions.
    H.K. was supported by JSPS KAKENHI Grants No.~JP23K25783 and No.~JP23K25790.
    K.F. was supported by JSPS KAKENHI Grant No.~JP25K23354.
    K.F. and H.K. were supported by MEXT KAKENHI Grant-in-Aid for Transformative Research Areas A "Extreme Universe" (KAKENHI Grant No.~JP21H05191).
\end{acknowledgments}

\section*{DATA AVAILABILITY}
No data were created or analyzed in this study.

\appendix

\onecolumngrid

\section{Proof of mutual commutativity of the transfer matrices}\label{app:transfer-mat-mutual-commutativity}
In this appendix, we prove the mutual commutativity of the transfer matrices in Eqs.~\eqref{eq:transfer-mat-mutual-commutativity} and~\eqref{eq:transfer-mat-mutual-commutativity-odd}.
We start from the RTT relation~\eqref{eq:RTT} and substitute the monodromy matrix in terms of the transfer matrices from Eq.~\eqref{eq:monodromy-matrix-fwd}.
Reversing the signs of $u$ and $v$, the monodromy matrix becomes $\monodromybwd{N}{a}{-u} = \gamma_a^{N} \qty(  \tmat{N}{+}{u} + \gamma_a \tmattil{N}{-}{u} )$ where $\tmattil{N}{-}{u} \equiv \sqrt{-\imi u} \tmat{N}{-}{u}$ and we evaluate both sides of the RTT relation.

For the left-hand side of Eq.~\eqref{eq:RTT}, we obtain
\begin{align}
      &
    R_{a,b}(\sqrt{u / v}) \monodromybwd{N}{a}{-u} \monodromybwd{N}{b}{-v}
    \nonumber  \\
    = &
    (\gamma_a - \sqrt{u/v} \gamma_b)
    \gamma_a^{N} \qty( \tmat{N}{+}{u} +\gamma_a \tmattil{N}{-}{u})
    \gamma_b^{N} \qty( \tmat{N}{+}{v} +\gamma_b \tmattil{N}{-}{v})
    \nonumber  \\
    = &
    (-\gamma_a \gamma_b)^{N}
    (\gamma_a - \sqrt{u/v} \gamma_b)
    \qty( \tmat{N}{+}{u} + (-1)^{N} \gamma_a \tmattil{N}{-}{u})
    \qty( \tmat{N}{+}{v} + \gamma_b \tmattil{N}{-}{v})
    \nonumber  \\
    = &
    (-\gamma_a \gamma_b)^{N}
    \biggl\{
    \gamma_a \qty(\tmat{N}{+}{u} \tmat{N}{+}{v} - (-1)^{N} \sqrt{\frac{u}{v}} \tmattil{N}{-}{u} \tmattil{N}{-}{v})
    -
    \gamma_b \qty(  \sqrt{\frac{u}{v}}\tmat{N}{+}{u} \tmat{N}{+}{v} + (-1)^{N} \tmattil{N}{-}{u} \tmattil{N}{-}{v})
    \nonumber  \\
      & \qquad
    +(-1)^{N} \tmattil{N}{-}{u} \tmat{N}{+}{v} - \sqrt{\frac{u}{v}} \tmat{N}{+}{u} \tmattil{N}{-}{v}
    +
    \gamma_a \gamma_b \qty( \tmat{N}{+}{u} \tmattil{N}{-}{v} + (-1)^{N} \sqrt{\frac{u}{v}} \tmattil{N}{-}{u} \tmat{N}{+}{v})
    \biggr\}
\end{align}

Similarly, for the right-hand side of Eq.~\eqref{eq:RTT}, we have
\begin{align}
      &
    \monodromybwd{N}{b}{-v} \monodromybwd{N}{a}{-u} R_{a,b}(\sqrt{u / v})
    \nonumber  \\
    = &
    \gamma_b^{N} \qty( \tmat{N}{+}{v} +\gamma_b \tmattil{N}{-}{v})
    \gamma_a^{N} \qty( \tmat{N}{+}{u} +\gamma_a \tmattil{N}{-}{u})
    (\gamma_a - \sqrt{u/v} \gamma_b)
    \nonumber  \\
    = &
    (-\gamma_a \gamma_b)^{N} \qty( \tmat{N}{+}{v} + (-1)^{N} \gamma_b \tmattil{N}{-}{v})
    \qty( \tmat{N}{+}{u} +\gamma_a \tmattil{N}{-}{u})
    (\gamma_a - \sqrt{u/v} \gamma_b)
    \nonumber  \\
    = &
    (-\gamma_a \gamma_b)^{N}
    \biggl\{
    \gamma_a \qty(\tmat{N}{+}{v} \tmat{N}{+}{u} - (-1)^{N} \sqrt{\frac{u}{v}} \tmattil{N}{-}{v} \tmattil{N}{-}{u})
    -
    \gamma_b \qty(  \sqrt{\frac{u}{v}}\tmat{N}{+}{v} \tmat{N}{+}{u} + (-1)^{N} \tmattil{N}{-}{v} \tmattil{N}{-}{u})
    \nonumber  \\
      & \qquad
    - \tmat{N}{+}{v} \tmattil{N}{-}{u} + (-1)^{N} \sqrt{\frac{u}{v}} \tmattil{N}{-}{v} \tmat{N}{+}{u}
    +
    \gamma_a \gamma_b \qty( (-1)^N \tmattil{N}{-}{v} \tmat{N}{+}{u} +  \sqrt{\frac{u}{v}} \tmat{N}{+}{v} \tmattil{N}{-}{u})
    \biggr\}
\end{align}

Equating the coefficients of each power of the auxiliary Majorana fermions on both sides, we obtain the following relations.
From the coefficient of $\gamma_a$, we have
\begin{align}
    \label{eq:RTT-1}
    \tmat{N}{+}{u} \tmat{N}{+}{v} - (-1)^{N} \sqrt{\frac{u}{v}} \tmattil{N}{-}{u} \tmattil{N}{-}{v}
     & =
    \tmat{N}{+}{v} \tmat{N}{+}{u} - (-1)^{N} \sqrt{\frac{u}{v}} \tmattil{N}{-}{v} \tmattil{N}{-}{u}
    \,.
\end{align}
From the coefficient of $\gamma_b$, we have
\begin{align}
    \label{eq:RTT-2}
    \sqrt{\frac{u}{v}}\tmat{N}{+}{u} \tmat{N}{+}{v} + (-1)^{N} \tmattil{N}{-}{u} \tmattil{N}{-}{v}
     & =
    \sqrt{\frac{u}{v}}\tmat{N}{+}{v} \tmat{N}{+}{u} + (-1)^{N} \tmattil{N}{-}{v} \tmattil{N}{-}{u}
    \,.
\end{align}
From Eqs.~\eqref{eq:RTT-1} and~\eqref{eq:RTT-2}, we immediately obtain the mutual commutativity:
\begin{align}
    \qty[\tmat{N}{+}{u},\tmat{N}{+}{v}] = 0 \quad \text{and} \quad \qty[\tmattil{N}{-}{u},\tmattil{N}{-}{v}] = 0\,.
\end{align}

This completes the proof of Eqs.~\eqref{eq:transfer-mat-mutual-commutativity} and~\eqref{eq:transfer-mat-mutual-commutativity-odd}.

\section{Proof of Eq.~\eqref{eq:monodromy-matrix-fwd}}\label{app:monodromy-matrix-fwd}

Here, we prove Eq.~\eqref{eq:monodromy-matrix-fwd} by induction.
Equation~\eqref{eq:monodromy-matrix-bwd} can be proved similarly. 

Throughout this Appendix, we make the $N$-dependence explicit by writing the SYK charges as $\syk{N}{p} = \sykN{N}{p}$, the transfer matrices as $\tmat{N}{\pm}{u} = \tmatN{N}{\pm}{u}$, and the monodromy matrix as $\monodromyfwd{N}{a}{u} = \monodromyfwdN{N}{a}{u}$.

We can see by inspection that the SYK charges satisfy the following recursion relations:
\begin{align}
    \label{eq:syk-recursion}
    \begin{aligned}
        \sykN{N}{2p} = \sykN{N-1}{2p} + \sykN{N-1}{2p-1} \gamma_N
        \,,
        \\
        \sykN{N}{2p+1} = \sykN{N-1}{2p+1} + \imi \sykN{N-1}{2p} \gamma_N
        \,.
    \end{aligned}
\end{align}
These relations lead to the recursion relations for the transfer matrices:
\begin{align}
    \begin{aligned}
        \tmatN{N}{+}{u}
         & =
        \tmatN{N-1}{+}{u} -\imi u \tmatN{N-1}{-}{u} \gamma_N
        \,,
        \\
        \tmatN{N}{-}{u}
         & =
        \tmatN{N-1}{-}{u}
        +
        \tmatN{N-1}{+}{u} \gamma_N
        \,.
    \end{aligned}
    \label{eq:transfer-mat-recursion}
\end{align}

Equation~\eqref{eq:monodromy-fwd} for the base case $N=1$ holds trivially.
Let us assume Eq.~\eqref{eq:monodromy-matrix-fwd} holds for $N = M-1$ case: $\monodromyfwdN{M-1}{a}{u} = \qty( \tmatN{M-1}{+}{u} - \gamma_a \sqrt{\imi u} \tmatN{M-1}{-}{u}) \gamma_a^{M-1}$.
Then we can calculate for the $N=M$ case as
\begin{align}
    \monodromyfwdN{M}{a}{u}
     & =
    \monodromyfwdN{M-1}{a}{u} R_{a,M}((-1)^{M} \sqrt{\imi u})
    \nonumber \\
     & =
    \qty( \tmatN{M-1}{+}{u} - \gamma_a \sqrt{\imi u} \tmatN{M-1}{-}{u}) \gamma_a^{M-1}
    (\gamma_a + (-1)^{M-1} \sqrt{\imi u} \gamma_M)
    \nonumber \\
     & =
    \qty( \tmatN{M-1}{+}{u} - \gamma_a \sqrt{\imi u} \tmatN{M-1}{-}{u})
    (1 + \sqrt{\imi u} \gamma_M \gamma_a) \gamma_a^{M}
    \nonumber \\
     & =
    \qty[ \tmatN{M-1}{+}{u} - \imi u \tmatN{M-1}{-}{u} \gamma_M
        - \gamma_a \sqrt{\imi u} \qty( \tmatN{M-1}{-}{u} + \tmatN{M-1}{+}{u} \gamma_M)  ]
    \gamma_a^{M}
    \nonumber \\
     & =
    \qty( \tmatN{M}{+}{u} - \gamma_a \sqrt{\imi u} \tmatN{M}{-}{u}) \gamma_a^{M}
    \,,
\end{align}
where in the second equality, we have used the assumption of the induction here, and in the last equality, we have used Eq.~\eqref{eq:transfer-mat-recursion}.
Then we have proved Eq.~\eqref{eq:monodromy-matrix-fwd}.

\section{Proof of Eq.~\eqref{eq:char-poly}}\label{app:char-poly}
Here we will prove Eq.~\eqref{eq:char-poly}.
From Eqs.~\eqref{eq:monodromy-matrix-fwd}-\eqref{eq:monodromy-inversion}, we can see
\begin{align}
    (1 + \imi u)^N
     & =
    \monodromyfwd{N}{a}{u} \monodromybwd{N}{a}{u}
    \nonumber \\
     & =
    \qty( \tmat{N}{+}{u} - \gamma_a \sqrt{\imi u} \tmat{N}{-}{u})
    \qty(  \tmat{N}{+}{-u} + \gamma_a \sqrt{\imi u} \tmat{N}{-}{-u} )
    \nonumber \\
     & =
    \tmat{N}{+}{u} \tmat{N}{+}{-u} + \imi u \tmat{N}{-}{u} \tmat{N}{-}{-u}
    +
    \gamma_a \sqrt{\imi u} \qty(\tmat{N}{+}{u} \tmat{N}{-}{-u} - \tmat{N}{-}{u} \tmat{N}{+}{-u} )
    \,,
    \label{eq:monodromy-prod-expand}
\end{align}
where in the second line, we have used Eqs.~\eqref{eq:monodromy-matrix-fwd} and~\eqref{eq:monodromy-matrix-bwd}.
The left-hand side does not have the auxiliary Majorana fermion $\gamma_a$, and the first term in the right-hand side does not have $\gamma_a$, the second term does, thus we can see
\begin{align}
    \label{eq:poly-tautau-relation}
    \tmat{N}{+}{u} \tmat{N}{+}{-u} + \imi u \tmat{N}{-}{u} \tmat{N}{-}{-u}
    =
    (1 + \imi u)^N
    \,,
\end{align}
and flipping the sign of $u$ in the above equation, we have
\begin{align}
    \label{eq:poly-tautau-relation-minus}
    \tmat{N}{+}{u} \tmat{N}{+}{-u} - \imi u \tmat{N}{-}{u} \tmat{N}{-}{-u}
    =
    (1 - \imi u)^N
    \,,
\end{align}
where we have used the mutual commutativity~\eqref{eq:transfer-mat-mutual-commutativity} and~\eqref{eq:transfer-mat-mutual-commutativity-odd}.
%
The second term in the right-hand side of Eq.~\eqref{eq:monodromy-prod-expand} has auxiliary Majorana $\gamma_a$, whose coefficient must be zero:
\begin{align}
    \tmat{N}{+}{u} \tmat{N}{-}{-u} = \tmat{N}{-}{u} \tmat{N}{+}{-u}
    \,.
\end{align}

From Eqs.~\eqref{eq:poly-tautau-relation} and~\eqref{eq:poly-tautau-relation-minus}, we have
\begin{align}
    \tmat{N}{+}{u} \tmat{N}{+}{-u}
     & =
    \frac{(1 + \imi u)^N + (1 - \imi u)^N}{2}
    \,,
    \\
    \tmat{N}{-}{u} \tmat{N}{-}{-u}
     & =
    \frac{(1 + \imi u)^N - (1 - \imi u)^N}{2 \imi u}
    \,.
\end{align}
Thus, we have completed the proof of Eq.~\eqref{eq:char-poly}.

\section{Proof of Eqs.~\eqref{eq:gamma1-conjugation-even} and~\eqref{eq:gamma1-conjugation-odd}}
\label{app:gamma1-conjugation}

Here, we prove Eqs.~\eqref{eq:gamma1-conjugation-even} and~\eqref{eq:gamma1-conjugation-odd}.

We first calculate the conjugation of Majorana fermions with the monodromy matrix.
We use the following relation:
\begin{align}
    R_{a,j}((-1)^{j} \sqrt{\imi u}) \gamma_l R_{a,j}((-1)^{j} \sqrt{\imi u})
    & =
    \begin{cases}
         - (1 + \imi u) \gamma_l & (l \notin \{j, a\})
         \\
         (-1 + \imi u) \gamma_j + 2 (-1)^{j} \sqrt{\imi u} \gamma_a & (l = j)
         \\
         (1 - \imi u) \gamma_a + 2 (-1)^{j} \sqrt{\imi u} \gamma_j & (l = a)
    \end{cases}
    \,.
\end{align}
Repeatedly using these relations, we obtain
\begin{align}
     &
    \monodromyfwd{N}{a}{u}
    \gamma_j
    \monodromybwd{N}{a}{u}
    =
     \qty(\prod_{j=1}^{N} R_{a,j}((-1)^{j} \sqrt{\imi u}))
     \gamma_j
     \qty(\prod_{j=N}^{1} R_{a,j}((-1)^{j} \sqrt{\imi u}))
    \nonumber       \\
     = &
    (-1 - \imi  u)^{N-1} (-1 + \imi u) \gamma_j
    -
    \sum_{l = 1}^{j-1} 4 \imi u (-1 - \imi  u)^{N+l-j-1} (-1 + \imi u)^{j-1-l} \gamma_{l}
    +
    2 (-1 - \imi  u)^{N-j} \sqrt{\imi u} (-1 + \imi u)^{j-1} \gamma_a
    \,.
            \label{eq:gamma-conjugate-fb}
\end{align}
Similarly, we also have
\begin{align}
     & \monodromybwd{N}{a}{-u}
    \gamma_j
    \monodromyfwd{N}{a}{-u}
    =
     \qty(\prod_{j=N}^{1} R_{a,j}((-1)^{j} \sqrt{-\imi u}))
     \gamma_j
     \qty(\prod_{j=1}^{N} R_{a,j}((-1)^{j} \sqrt{-\imi u}))
    \nonumber                 \\
     =&
    (-1 + \imi  u)^{N-1} (-1 - \imi u) \gamma_j
    -
    \sum_{l = j+1}^{N} 4 \imi u (-1 + \imi  u)^{N-l+j-1} (-1 - \imi u)^{l-j-1} \gamma_{l}
    +
    2 (1 - \imi  u)^{j-1} \sqrt{\imi u} (1 + \imi u)^{N-j} \gamma_a
    \,.
        \label{eq:gamma-conjugate-bf}
\end{align}

Meanwhile, from Eqs.~\eqref{eq:monodromy-matrix-fwd} and~\eqref{eq:monodromy-matrix-bwd}, we have
\begin{align}
     &
    \monodromyfwd{N}{a}{u}
    \gamma_j
    \monodromybwd{N}{a}{u}
    \nonumber \\
    & =
    \qty( \tmat{N}{+}{u} - \gamma_a \sqrt{\imi u} \tmat{N}{-}{u}) \gamma_a^{N}
    \gamma_j
    \gamma_a^{N} \qty( \tmat{N}{+}{-u} + \gamma_a \sqrt{\imi u} \tmat{N}{-}{-u} )
    \nonumber \\
     & =
    (-1)^N
    \qty{
        \tmat{N}{+}{u} \gamma_j \tmat{N}{+}{-u}
        - \imi u
        \tmat{N}{-}{u} \gamma_j \tmat{N}{-}{-u}
    }
    +
    (-1)^{N+1} \sqrt{\imi u}
    \gamma_a
    \qty{
        \tmat{N}{-}{u} \gamma_j \tmat{N}{+}{-u}
        +
        \tmat{N}{+}{u} \gamma_j \tmat{N}{-}{-u}
    }
    \,,
        \label{eq:gamma-conjugate-fb-in-tau}
    \\
     &
    \monodromybwd{N}{a}{-u}
    \gamma_j
    \monodromyfwd{N}{a}{-u}
    \nonumber \\
    & =
    \gamma_a^{N} \qty( \tmat{N}{+}{u} + \gamma_a \sqrt{-\imi u} \tmat{N}{-}{u} )
    \gamma_j
    \qty( \tmat{N}{+}{-u} - \gamma_a \sqrt{-\imi u} \tmat{N}{-}{-u}) \gamma_a^{N}
    \nonumber \\
     & =
    (-1)^{N}
    \qty{
        \tmat{N}{+}{u} \gamma_j \tmat{N}{+}{-u}
        + \imi u
        \tmat{N}{-}{u} \gamma_j \tmat{N}{-}{-u}
    }
    +
    \sqrt{-\imi u}
    \gamma_a
    \qty{
        \tmat{N}{-}{u} \gamma_j \tmat{N}{+}{-u}
        +
        \tmat{N}{+}{u} \gamma_j \tmat{N}{-}{-u}
    }
    \,.
    \label{eq:gamma-conjugate-bf-in-tau}
\end{align}

Comparing Eqs.~\eqref{eq:gamma-conjugate-fb} and~\eqref{eq:gamma-conjugate-fb-in-tau}, and also comparing Eqs.~\eqref{eq:gamma-conjugate-bf} and~\eqref{eq:gamma-conjugate-bf-in-tau}, we obtain
\begin{align}
        \tmat{N}{+}{u} \gamma_j \tmat{N}{+}{-u}
        - \imi u
        \tmat{N}{-}{u} \gamma_j \tmat{N}{-}{-u}
     & =
    (1 + \imi  u)^{N-1} (1 - \imi u) \gamma_j
    -
    \sum_{l = 1}^{j-1} 4 \imi u (1 + \imi  u)^{N+l-j-1} (1 - \imi u)^{j-1-l} \gamma_{l}
    \,,
    \\
        \tmat{N}{+}{u} \gamma_j \tmat{N}{+}{-u}
        + \imi u
        \tmat{N}{-}{u} \gamma_j \tmat{N}{-}{-u}
     & =
    (1 - \imi  u)^{N-1} (1 + \imi u) \gamma_j
    +
    \sum_{l = j+1}^{N} 4 \imi u (1 - \imi  u)^{N-l+j-1} (1 + \imi u)^{l-j-1} \gamma_{l}
    \,,
\end{align}
and hence
\begin{align}
    \tmat{N}{+}{u} \gamma_j \tmat{N}{+}{-u}
     & =
    \frac{1}{2}
    \qty((1 + \imi  u)^{N-1} (1 - \imi u) + (1 - \imi  u)^{N-1} (1 + \imi u)) \gamma_j
    \nonumber \\
     & \quad
    -
    \sum_{l = 1}^{j-1} 2 \imi u (1 + \imi  u)^{N+l-j-1} (1 - \imi u)^{j-1-l} \gamma_{l}
    +
    \sum_{l = j+1}^{N} 2 \imi u (1 - \imi  u)^{N-l+j-1} (1 + \imi u)^{l-j-1} \gamma_{l}
    \,,
    \\
    \tmat{N}{-}{u} \gamma_j \tmat{N}{-}{-u}
     & =
    -
    \frac{1}{2 \imi u}
    \qty((1 + \imi  u)^{N-1} (1 - \imi u) - (1 - \imi  u)^{N-1} (1 + \imi u)) \gamma_j
    \nonumber \\
     & \quad
    +
    \sum_{l = 1}^{j-1} 2 (1 + \imi  u)^{N+l-j-1} (1 - \imi u)^{j-1-l} \gamma_{l}
    +
    \sum_{l = j+1}^{N} 2 (1 - \imi  u)^{N-l+j-1} (1 + \imi u)^{l-j-1} \gamma_{l}
    \,.
\end{align}

Through the variable transformation $u = \tan(\kappa / 2)$, we have
\begin{align}
    1 \pm \imi u
    =
    \frac{e^{\pm \imi \kappa / 2}}{\cos(\kappa/2)}
    \,,
\end{align}
and the expressions become
\begin{align}
     \tmat{N}{+}{u} \gamma_j \tmat{N}{+}{-u}
     & =
    \frac{\cos((N-2)\kappa/2)}{\cos^{N}(\kappa/2)}
    \gamma_j
    -
    \sum_{l = 1}^{j-1} 2 \imi \frac{\sin(\kappa/2)e^{\imi (N/2+(l-j)) \kappa}}{\cos^{N-1}(\kappa/2)} \gamma_{l}
    +
    \sum_{l = j+1}^{N} 2 \imi \frac{\sin(\kappa/2)e^{-\imi (N/2-(l-j)) \kappa}}{\cos^{N-1}(\kappa/2)} \gamma_{l}
    \,,
    \label{eq:gammaj-conjugation-even}
    \\
     \tmat{N}{-}{u} \gamma_j \tmat{N}{-}{-u}
     & =
    - \frac{\sin((N-2)\kappa/2)}{\sin(\kappa/2) \cos^{N-1}(\kappa/2)}
    \gamma_j
    +
    2
    \sum_{l = 1}^{j-1} \frac{e^{\imi (N/2+(l-j)) \kappa}}{\cos^{N-2}(\kappa/2)} \gamma_{l}
    +
    2
    \sum_{l = j+1}^{N} \frac{e^{-\imi (N/2-(l-j)) \kappa}}{\cos^{N-2}(\kappa/2)} \gamma_{l}
    \,.
    \label{eq:gammaj-conjugation-odd}
\end{align}
Substituting $j = 1$ into these expressions, we obtain Eqs.~\eqref{eq:gamma1-conjugation-even} and~\eqref{eq:gamma1-conjugation-odd}, completing the proof.

\section{Proof of derivation of the critical Ising Hamiltonian}\label{app:log-derivative-ising}
Here, we prove Eq.~\eqref{eq:log-derivative-ising}, which relates the logarithmic derivative of the transfer matrix to the critical Ising Hamiltonian.
We also prove Eq.~\eqref{eq:ising-in-fermion-mode}.

Majorana fermions are expressed 
in terms of complex fermions $f_k$ as
\begin{align}
    \label{eq:fermion-inverse}
    \gamma_j
     & =
    \sqrt{\frac{2}{N}} \sum_{k \in \modeset{N}{\pm}} e^{-\imi (j-1)k } f_k
    \,,
\end{align}
where $\modeset{N}{+} = \left\{ k \in \frac{2 \pi}{N} \qty(\mathbb{Z} + \frac{1}{2}) : -\pi < k\leq \pi \right\}$ and $\modeset{N}{-} = \left\{ k \in \frac{2 \pi}{N} \mathbb{Z} : -\pi < k\leq \pi \right\}$, $f_{-k} \equiv f_k^\dag$, and we defined the zero-momentum mode $f_0 \equiv \frac{1}{\sqrt{2}}\chi_0$ and the $\pi$-momentum mode $f_\pi \equiv \frac{1}{\sqrt{2}}\chi_\pi$.

Then, we rewrite the Ising Hamiltonian in the complex fermions:
\begin{align}
    \isingH{N}{\pm}
    &=
    \imi
    \sum_{j=1}^{N-1}
    \gamma_j \gamma_{j+1}
    \pm
    \imi
    \gamma_{N} \gamma_{1}
    =
    \frac{2\imi}{N}
    \sum_{k,l \in \modeset{N}{\mp}}
        \qty(\sum_{j=1}^{N}e^{-\imi (j-1)(k+l) })
        e^{\imi l } f_k f_l
    \nonumber\\
    &=
    4
    \sum_{k \in \modesetpositive{N}{\mp}}
    \sin k  ( f_k^\dag f_k - 1/2)
    \,.
    \label{eq:Ising-in-f-proof}
\end{align}

The logarithmic derivative of the transfer matrix~\eqref{eq:tau-fermion-prod} becomes
\begin{align}
    \pdv{u} \eval{\log \tmat{N}{\pm}{u}}_{u=-\imi}
     & =
    \sum_{k \in \modesetpositive{N}{\pm}}
    \pdv{u} \eval{\log(1 - u \epsilon_k (n_k - 1/2))}_{u=-\imi}
    \,.
\end{align}
The right-hand side is calculated as
\begin{align}
     \sum_{k \in \modesetpositive{N}{\pm}}
    \pdv{u} \eval{\log(1 - u \epsilon_k (n_k - 1/2))}_{u=-\imi}
     & =
    -\sum_{k \in \modesetpositive{N}{\pm}}
    \frac{\epsilon_k (n_k - 1/2)}{1 + \imi \epsilon_k (n_k - 1/2)}
    \nonumber                                \\
     & =
    -\sum_{k \in \modesetpositive{N}{\pm}}
    \frac{\epsilon_k}{1 + \epsilon_k^2 / 4}
    ((n_k - 1/2) - \imi \epsilon_k / 4)
    \nonumber                                \\
     & =
    -
    \sum_{k \in \modesetpositive{N}{\pm}}
    \sin k
    (n_k - 1/2)
    +
    \imi
    \sum_{k \in \modesetpositive{N}{\pm}}
    \cos^2( k/2)
    \,,
    \nonumber                                \\
     & =
    - \frac{1}{4} \isingH{N}{\mp}
    +
    \frac{\imi}{4}(N \pm 1 -1)
    \,,
    \label{eq:log-derivative-mid}
\end{align}
where we used $\epsilon_k = 2 \cot( k/2)$, and in the last equality, we used Eq.~\eqref{eq:Ising-in-f-proof} and the following relation for the constant term:
\begin{align}
    \imi
    \sum_{k \in \modesetpositive{N}{\pm}}
    \cos^2( k/2)
    =
    \frac{\imi}{2}
    \sum_{k \in \modesetpositive{N}{\pm}}
    [\cos k + 1]
    =
    \frac{\imi}{4}(N \pm 1 -1)
    \,.
\end{align}

Thus, we have completed the proof of Eq.~\eqref{eq:log-derivative-ising}.
Equation~\eqref{eq:Ising-in-f-proof} also establishes Eq.~\eqref{eq:ising-in-fermion-mode}.

\section{Proof of Eq.~\eqref{eq:fermion-FFD-like}}
\label{app:fermion-FFD-like}

Here we prove Eq.~\eqref{eq:fermion-FFD-like}.
Substituting $u = u_k = \tan(k/2)$ for $k \in \modesetpositive{N}{+}$ into Eq.~\eqref{eq:gammaj-conjugation-even}, we have
\begin{align}
     \tmat{N}{+}{u_k} \gamma_j \tmat{N}{+}{-u_k}
     & =
    \frac{\cos((N-2)k/2)}{\cos^{N}(k/2)}
    \gamma_j
    -
    \sum_{l = 1}^{j-1} 2 \imi \frac{\sin(k/2)e^{\imi (N/2+(l-j)) k}}{\cos^{N-1}(k/2)} \gamma_{l}
    +
    \sum_{l = j+1}^{N} 2 \imi \frac{\sin(k/2)e^{-\imi (N/2-(l-j)) k}}{\cos^{N-1}(k/2)} \gamma_{l}
    \nonumber\\
    &=
    \qty(
    \frac{\sin(Nk/2) \sin(k/2)}{\cos^{N}(k/2)}
    -
    2 \imi \frac{\sin(k/2)e^{-\imi (N/2) k}}{\cos^{N-1}(k/2)}
    ) \gamma_{j}
    +
    \sum_{l = 1 }^{N} 2 \imi \frac{\sin(k/2)e^{-\imi (N/2-(l-j)) k}}{\cos^{N-1}(k/2)} \gamma_{l}
    \nonumber\\
    &=
    \sqrt{8N} \frac{\sin(k/2)\sin(Nk/2)}{\cos^{N-1}(k/2)}
    e^{-\imi (j-1) k}
    f_k
    \,,
\end{align}
where we used the relation $e^{\imi Nk/2} = \imi \sin(Nk/2)$ for $k \in \modeset{N}{+}$.

In the same way, substituting $u = u_k = \tan(k/2)$ for $k \in \modesetpositive{N}{-}$ into Eq.~\eqref{eq:gammaj-conjugation-odd}, we have
\begin{align}
      \tmat{N}{-}{u_k} \gamma_j \tmat{N}{-}{-u_k}
     & =
    - \frac{\sin((N-2)k/2)}{\sin(k/2) \cos^{N-1}(k/2)}
    \gamma_j
    +
    2
    \sum_{l = 1}^{j-1} \frac{e^{\imi (N/2+(l-j)) k}}{\cos^{N-2}(k/2)} \gamma_{l}
    +
    2
    \sum_{l = j+1}^{N} \frac{e^{-\imi (N/2-(l-j)) k}}{\cos^{N-2}(k/2)} \gamma_{l}
    \nonumber\\
    &=
    \qty(
        \frac{\cos(Nk/2) \sin(k)}{\sin(k/2) \cos^{N-1}(k/2)} - \frac{2e^{-\imi Nk/2}}{\cos^{N-2}(k/2)}
    )
    \gamma_j
    +
    2
    \sum_{l = 1}^{N} \frac{e^{-\imi (N/2-(l-j)) k}}{\cos^{N-2}(k/2)} \gamma_{l}
    \nonumber\\
    &=
    \sqrt{8N}
    \frac{\cos(Nk/2)}{\cos^{N-2}(k/2)} e^{-\imi(j-1)k} f_k
    \,.
\end{align}

Summarizing the above results, we obtain
\begin{align}
    \tmat{N}{\pm}{u_k} \gamma_j \tmat{N}{\pm}{-u_k}
     & =
    \mathcal{N}_{k} e^{-\imi  (j-1) k} f_k
    \quad (k \in \modesetpositive{N}{\pm})
    \,,
\end{align}
where the normalization factor is given by
\begin{align}
    \label{eq:normalization-factor-simple}
    \mathcal{N}_{k} =
    \begin{cases}
        \displaystyle
        \sqrt{8N} \frac{\sin(k/2)\sin(Nk/2)}{\cos^{N-1}(k/2)}
        & (k \in \modesetpositive{N}{+})
        \,,
        \\[1ex]
        \displaystyle
        \sqrt{8N} \frac{\cos(Nk/2)}{\cos^{N-2}(k/2)}
        & (k \in \modesetpositive{N}{-})
        \,.
    \end{cases}
\end{align}
We note that $\sin(Nk/2) = \pm 1$ for $k \in \modeset{N}{+}$ and $\cos(Nk/2) = \pm 1$ for $k \in \modeset{N}{-}$.
This concludes the proof of Eq.~\eqref{eq:fermion-FFD-like}.

Below, we express the normalization factor~\eqref{eq:normalization-factor-simple} using the characteristic polynomial~\eqref{eq:char-poly}.
In the following, we often use the notation for the derivative of the polynomials as~\footnote{In~\cite{disguise-Fendley-2019}, the notation $P'_{N}(u_k^2)$ is used for $\Dpoly{N}{\pm}{u_k^2}$ in this paper.}
\begin{align}
    \Dpoly{N}{\pm}{u_k^2} & \equiv \pdv{u}\eval{\poly{N}{\pm}{u^2}}_{u=u_k}
    \quad (k \in \modesetpositive{N}{\pm})
    \,,
\end{align}
where $u_k = \tan(k/2)$.
The explicit expression is given by
\begin{align}
    \Dpoly{N}{\pm}{u_k^2}
    =
    \imi N \frac{(1+\imi u_k)^{N-1} \mp (1-\imi u_k)^{N-1}}{(1+\imi u_k) \pm (1-\imi u_k)}
    \quad (k \in \modesetpositive{N}{\pm})
    \,,
\end{align}
and we can further simplify this expression as
\begin{align}
    \Dpoly{N}{+}{u_k^2}
    &=
    - N \frac{\sin(Nk/2)}{\cos^{N-2}(k/2)}
    \quad (k \in \modesetpositive{N}{+})
    \,,
    \\
    \Dpoly{N}{-}{u_k^2}
    &=
    N \frac{\cos(Nk/2)}{\cos^{N-3}(k/2)\sin(k/2)}
    \quad (k \in \modesetpositive{N}{-})
    \,.
\end{align}
Then, the normalization factor is expressed as
\begin{align}
    \label{eq:normalization-factor}
    \mathcal{N}_{k} = \mp \sqrt{\frac{8}{N}} u_k \Dpoly{N}{\pm}{u_k^2}
    \quad (k \in \modesetpositive{N}{\pm})
    \,.
\end{align}

We note that the normalization factor has the symmetry
\begin{align}
    \label{eq:normalization-factor-symm}
    \mathcal{N}_{k} =  \mathcal{N}_{-k}
    \quad (k \in \modesetpositive{N}{\pm})
    \,.
\end{align}

\section{Proof of Eq.~\eqref{eq:tau-fermion-relation}}\label{app:tau-fermion-relation}

In this Appendix, we prove Eq.~\eqref{eq:tau-fermion-relation} in two different ways.
One utilizes the monodromy matrices~\eqref{eq:monodromy-fwd} and~\eqref{eq:monodromy-bwd}, and the other is based on the calculation of commutators and anticommutators of the SYK charges and the complex fermion modes.
Both strategies yield the same result.

\subsection{Proof of Eq.~\eqref{eq:tau-fermion-relation} using monodromy matrices}

Using Eq.~\eqref{eq:fermion-FFD-like}, we have
\begin{align}
    \tmat{N}{\pm}{u} f_k \tmat{N}{\pm}{-u}
    & =
    \frac{1}{\mathcal{N}_{k}}
    \tmat{N}{\pm}{u} \tmat{N}{\pm}{u_k} \gamma_1 \tmat{N}{\pm}{-u_k} \tmat{N}{\pm}{-u}
    \nonumber\\
    & =
    \frac{1}{\mathcal{N}_{k}}
    \tmat{N}{\pm}{u_k} \qty(\tmat{N}{\pm}{u} \gamma_1 \tmat{N}{\pm}{-u}) \tmat{N}{\pm}{-u_k}
    \quad(k \in \modesetpositive{N}{\pm})
    \,.
\end{align}
For $k \in \modesetpositive{N}{+}$, using Eq.~\eqref{eq:gammaj-conjugation-even}, we have
\begin{align}
    \tmat{N}{+}{u} f_k \tmat{N}{+}{-u}
    &=
    \frac{1}{\mathcal{N}_{k}}
    \tmat{N}{\pm}{u_k} \qty[
        \frac{\cos((N/2-1)\kappa)}{\cos^{N}(\kappa/2)}
        \gamma_1
        +
        \sum_{l = 2}^{N} 2 \imi \frac{\sin(\kappa/2)e^{-\imi (N/2-(l-1)) \kappa}}{\cos^{N-1}(\kappa/2)} \gamma_{l}
    ] \tmat{N}{\pm}{-u_k}
    \nonumber\\
    &=
     \frac{\cos((N/2-1)\kappa)}{\cos^{N}(\kappa/2)} f_k
        +
    \imi
    \frac{e^{-\imi N \kappa / 2} \sin\kappa}{\cos^{N}(\kappa/2)}
    \qty(\sum_{l = 2}^{N} e^{\imi (l-1) (\kappa - k)}) f_k
    \nonumber\\
    &=
    \frac{\cos(N\kappa/2)}{\cos^{N}(\kappa/2)} \frac{\sin((k + \kappa)/2)}{\sin((k - \kappa)/2)}
    f_k
    \nonumber\\
    &=
    \poly{N}{+}{u^2}
    \frac{u_k + u}{u_k - u}
    f_k
    \,.
\end{align}
For $k \in \modesetpositive{N}{-}$, using Eq.~\eqref{eq:gammaj-conjugation-odd}, we have
\begin{align}
    \tmat{N}{-}{u} f_k \tmat{N}{-}{-u}
    &=
    \frac{1}{\mathcal{N}_{k}}
    \tmat{N}{\pm}{u_k}
    \qty[
        - \frac{\sin((N-2)\kappa/2)}{\sin(\kappa/2) \cos^{N-1}(\kappa/2)}
        \gamma_1
        +
        2
        \sum_{l = 2}^{N} \frac{e^{-\imi (N/2- (l-1)) \kappa}}{\cos^{N-2}(\kappa/2)} \gamma_{l}
    ]
    \tmat{N}{\pm}{-u_k}
    \nonumber\\
    &=
    - \frac{\sin((N-2)\kappa/2)}{\sin(\kappa/2) \cos^{N-1}(\kappa/2)}
    f_k
    +
    \frac{2 e^{-\imi N\kappa/2}}{\cos^{N-2}(\kappa/2)}
    \qty( \sum_{l = 2}^{N} e^{\imi (l-1)(\kappa-k)} ) f_k
   \nonumber\\
    &=
    -
    \frac{\sin(N\kappa/2)}{\sin(\kappa/2) \cos^{N-1}(\kappa/2)}
    \frac{\sin((k + \kappa)/2)}{\sin((k - \kappa)/2)}
   f_k
   \nonumber\\
    &=
    -
    \poly{N}{-}{u^2}
    \frac{u_k + u}{u_k - u}
    f_k
    \,.
\end{align}
Summarizing the above results, we have
\begin{align}
    \tmat{N}{\pm}{u} f_k \tmat{N}{\pm}{-u}
    =
    \pm
    \poly{N}{\pm}{u^2}
    \frac{u_k + u}{u_k - u}
    f_k
    \quad(k \in \modesetpositive{N}{\pm})
    \,.
\end{align}
Multiplying both sides by $\tmat{N}{\pm}{u}$ from the right, we obtain Eq.~\eqref{eq:tau-fermion-relation}.
This concludes the proof.

\subsection{Alternative proof of Eq.~\eqref{eq:tau-fermion-relation}}
We provide an alternative proof of Eq.~\eqref{eq:tau-fermion-relation}, which is rewritten here for convenience:
\begin{align}
    u_k \qty[\tmat{N}{\pm}{u}, f_k]_{\mp}
    =
    u \qty[\tmat{N}{\pm}{u}, f_k]_{\pm}
    \quad
    (k \in \modesetpositive{N}{\pm})
    \,,
\end{align}
where we define the notation for the commutator and anticommutator as
\begin{align}
    \qty[X, Y]_{\pm} & \equiv XY \pm YX
    \,,
\end{align}
and comparing the coefficients of $u^{p}$ on both sides, we have ($p \ge 0$)
\begin{align}
    \label{eq:H-f-commu-relation}
    \qty[\syk{N}{p+2}, f_k]_{(-)^{p+1}} = \varepsilon_k \qty[\syk{N}{p}, f_k]_{(-)^{p}}
    \quad
    (k \in \modesetpositive{N}{(-)^p})
    \,,
\end{align}
where we have used $\varepsilon_k = 1 / u_k  = \epsilon_k / 2 = \cot( k/2)$.
Below, we will prove Eq.~\eqref{eq:H-f-commu-relation}.

Then we can see the commutator and anticommutator of $\syk{N}{p}$ and $\gamma_j$ as follows.
\begin{align}
    \qty[\syk{N}{p}, \gamma_j]_{(-)^{p+1}}
     & = \imi^{\floor{p/2}}\sum_{1 \le i_1 < \cdots < i_{p} \le N} \qty[\gamma_{i_1} \cdots \gamma_{i_{p}}, \gamma_j]_{(-)^{p+1}}
    \nonumber                                                                                                                     \\
     & = (-1)^{p+1} 2 \imi^{\floor{p/2}}
    \sum_{l=0}^{p-1}
    (-1)^{l}
    \sum_{1 \le i_1 <\cdots <i_{l} < j }
    \sum_{j < i_{l+1} < \cdots <i_{p-1} \le N }
    \gamma_{i_1} \cdots \gamma_{i_{l}}
    \gamma_{i_{l+1}} \cdots \gamma_{i_{p-1}}
    \,,
    \\
    \qty[\syk{N}{p}, \gamma_j]_{(-)^{p}}
     & =
    \imi^{\floor{p/2}}
    \sum_{1 \le i_1 < \cdots < i_{p} \le N}
    \qty[\gamma_{i_1} \cdots \gamma_{i_{p}}, \gamma_j]_{(-)^{p}}
    \nonumber                                                                                                                     \\
     & =
    (-1)^{p} 2 \imi^{\floor{p/2}}
    \sum_{l=0}^{p}
    (-1)^{l}
    \sum_{1 \le i_1 <\cdots <i_{l} < j }
    \sum_{j < i_{l+1} < \cdots <i_{p} \le N }
    \gamma_{i_1} \cdots \gamma_{i_{l}}
    \gamma_j
    \gamma_{i_{l+1}} \cdots \gamma_{i_{p}}
    \,.
    \label{eq:H-gamma-comm2}
\end{align}

In the following, the mode $k$ satisfies $k \in \modesetpositive{N}{(-)^p}$.
The commutator of the SYK charges and the complex fermion is calculated as
\begin{align}
    \qty[\syk{N}{p}, f_k]_{(-)^{p+1}}
    = &
    \frac{\imi^{\floor{p/2}}}{\sqrt{2N}}\sum_{j=1}^{N} 
    e^{\imi (j-1) k}
    \sum_{1 \le i_1 < \cdots < i_{p} \le N} \qty[\gamma_{i_1} \cdots \gamma_{i_{p}}, \gamma_j]_{(-)^{p+1}}
    \nonumber \\
    = &
    (-1)^{p} \imi^{\floor{p/2}} \sqrt{\frac{2}{N}} \sum_{j=1}^{N} e^{\imi  (j-1) k}
    \sum_{l=0}^{p-1}
    (-1)^{l}
    \sum_{1 \le i_1 <\cdots <i_{l} < j }
    \sum_{j < i_{l+1} < \cdots <i_{p-1} \le N }
    \gamma_{i_1} \cdots \gamma_{i_{l}}
    \gamma_{i_{l+1}} \cdots \gamma_{i_{p-1}}
    \nonumber \\
    = &
    (-1)^{p} \imi^{\floor{p/2}} \sqrt{\frac{2}{N}}
    \sum_{1 \le i_1 <\cdots <i_{p-1} \le N }
    \qty(
    \sum_{l=0}^{p-1}
    (-1)^{l}
    \sum_{i_l < j < i_{l+1}} e^{\imi  (j-1) k}
    )
    \gamma_{i_1} \cdots \gamma_{i_{p-1}}
    \nonumber \\
    = &
    (-1)^{p} \imi^{\floor{p/2}+1} \sqrt{\frac{2}{N}}
    \varepsilon_k
    \sum_{1 \le i_1 <\cdots <i_{p-1} \le N }
    \qty(
    \sum_{l=1}^{p-1}
    (-1)^{l}
    e^{\imi  (i_{l} - 1) k}
    )
    \gamma_{i_1} \cdots \gamma_{i_{p-1}}
    \label{eq:H-f-acom}
    \,,
\end{align}
where in the third line, we define $i_0 = 0$ and $i_{p} = N+1$, and in the last line, we have used the following summation formula:
\begin{align}
    \sum_{l=0}^{p-1}
    (-1)^{l}
    \sum_{i_l < j < i_{l+1}} e^{\imi  (j-1) k}
    = &
    \sum_{l=0}^{p-1}
    (-1)^{l}
    \frac{e^{\imi  i_l k} - e^{\imi  (i_{l+1} - 1) k}}{1-e^{\imi  k}}
    \nonumber \\
    = &
    \frac{1}{1-e^{\imi  k}}
    \sum_{l=0}^{p-1}
    (-1)^{l}
    \qty[e^{\imi  i_l k} - e^{\imi  (i_{l+1} - 1) k}]
    \nonumber \\
    = &
    \frac{1}{1-e^{\imi  k}}
    \qty{
        \sum_{l=1}^{p-1}
        (-1)^{l}
        e^{\imi  (i_{l} - 1) k} (1 + e^{\imi  k})
    }
    \nonumber \\
    = &
    \imi \cot( k / 2)
    \sum_{l=1}^{p-1}
    (-1)^{l}
    e^{\imi  (i_{l} - 1) k}
    \,.
\end{align}
Also, another commutator and anticommutator become
\begin{align}
    \qty[\syk{N}{p}, f_k]_{(-)^{p}}
    = &
    \frac{\imi^{\floor{p/2}}}{\sqrt{2N}}
    \sum_{j=1}^{N} e^{\imi  (j - 1) k}
    \sum_{1 \le i_1 < \cdots < i_{p} \le N} \qty[\gamma_{i_1} \cdots \gamma_{i_{p}}, \gamma_j]_{(-)^{p}}
    \nonumber \\
    = &
    (-1)^{p} \imi^{\floor{p/2}} \sqrt{\frac{2}{N}}
    \sum_{j=1}^{N} e^{\imi  (j - 1) k}
    \sum_{l=0}^{p}
    (-1)^{l}
    \sum_{1 \le i_1 <\cdots <i_{l} < j }
    \sum_{j < i_{l+1} < \cdots <i_{p} \le N }
    \gamma_{i_1} \cdots \gamma_{i_{l}}
    \gamma_j
    \gamma_{i_{l+1}} \cdots \gamma_{i_{p}}
    \nonumber \\
    = &
    (-1)^{p+1} \imi^{\floor{p/2}} \sqrt{\frac{2}{N}}
    \sum_{1 \le i_1 < \cdots <i_{p+1} \le N }
    \qty(
    \sum_{l=1}^{p+1}
    (-1)^{l}
    e^{\imi  (i_l - 1) k}
    )
    \gamma_{i_1} \cdots  \gamma_{i_{p+1}}
    \label{eq:H-f-acom2}
    \,.
\end{align}
From Eqs.~\eqref{eq:H-f-acom} and~\eqref{eq:H-f-acom2}, we can see that Eq.~\eqref{eq:H-f-commu-relation} holds.
This concludes the proof of Eq.~\eqref{eq:tau-fermion-relation}.
Also, using~\eqref{eq:H-gamma-comm2}, we can prove Eq.~\eqref{eq:even-odd-relation}.

\section{Proof of Eqs.~\eqref{eq:tau-fermion-prod} and~\eqref{eq:tau-fermion-prod-odd}}\label{app:tau-fermion-prod}
In this appendix, we prove Eqs.~\eqref{eq:tau-fermion-prod} and~\eqref{eq:tau-fermion-prod-odd}, which express the transfer matrix in terms of the complex fermions.
We follow a similar argument as in~\cite{disguise-Fendley-2019}.

We first define the operator:
\begin{align}
    s_k & \equiv 2 n_k - 1 = [f_k^\dag, f_{k}]
    \quad (k \in \modesetpositive{N}{\pm})
    \,.
\end{align}
We note that $s_k$ appears here as an operator whose eigenvalues are $\pm 1$, whereas in the main text we used the same symbol to denote the eigenvalue itself.

From Eqs.~\eqref{eq:tau-fermion-relation} and~\eqref{eq:char-poly}, we can see that the conjugation of the complex fermion with the transfer matrix is
\begin{align}
    \label{eq:fermion-conjugate}
    \tmat{N}{\pm}{u} f_k \tmat{N}{\pm}{-u}
    =
    \pm
    (u_k + u)\frac{\poly{N}{\pm}{u^2}}{u_k - u} f_k
    \quad (k \in \modesetpositive{N}{\pm})
    \,.
\end{align}
Using the inversion of the complex fermion~\eqref{eq:fermion-inverse} and Eq.~\eqref{eq:tau-fermion-relation}, we have
\begin{align}
    \label{eq:fermion-FFD-like-gen}
    \tmat{N}{\pm}{u} \gamma_j \tmat{N}{\pm}{-u}
    =
    \pm
    \sqrt{\frac{2}{N}}
    \sum_{k \in \modeset{N}{\pm}} e^{-\imi  (j-1) k}
    (u_k + u)\frac{\poly{N}{\pm}{u^2}}{u_k - u} f_k
    \quad (k \in \modesetpositive{N}{\pm})
    \,.
\end{align}
Setting $j=1$ in Eq.~\eqref{eq:fermion-FFD-like-gen}, and for $k \in \modesetpositive{N}{\pm}$, we have
\begin{align}
    s_k
     & =
    \frac{1}{\mathcal{N}_{k}}
    \lim_{u \rightarrow u_k} \qty[f_k^{\dag}, \tmat{N}{\pm}{u} \gamma_1 \tmat{N}{\pm}{-u}]
    \nonumber \\
     & =
     \pm
    \frac{1}{\mathcal{N}_{k}}
    \lim_{u \rightarrow u_k} \frac{u_k + u}{u_k - u} \tmat{N}{\pm}{u} \qty[f_k^{\dag}, \gamma_1] \tmat{N}{\pm}{-u}
    \nonumber \\
     & =
     \mp
    \frac{2u_k}{\mathcal{N}_{k}}
    \pdv{u}\eval{\qty(\tmat{N}{\pm}{u} \qty[f_k^{\dag}, \gamma_1] \tmat{N}{\pm}{-u})}_{u=u_k}
    \nonumber \\
     & =
     \mp
    \frac{2u_k}{\mathcal{N}_{k}}
    \qty(\Dtmat{N}{\pm}{u_k} \qty[f_k^{\dag}, \gamma_1] \tmat{N}{\pm}{-u_k} - \tmat{N}{\pm}{u_k} 
    \qty[f_k^{\dag}, \gamma_1]
    \Dtmat{N}{\pm}{-u_k})
    \nonumber \\
     & =
     \mp
    \frac{2u_k}{\mathcal{N}_{k}}
    \qty(\Dtmat{N}{\pm}{u_k} f_k^{\dag} \gamma_1 \tmat{N}{\pm}{-u_k} + \tmat{N}{\pm}{u_k} \gamma_1 f_k^{\dag} \Dtmat{N}{\pm}{-u_k})
    \nonumber \\
     & =
     \mp
    \frac{2u_k}{\mathcal{N}_{k}} \sqrt{\frac{2}{N}} \qty(\Dtmat{N}{\pm}{u_k} \tmat{N}{\pm}{-u_k} + \tmat{N}{\pm}{u_k} \Dtmat{N}{\pm}{-u_k})
    \nonumber \\
     & =
    \frac{1}{\Dpoly{N}{\pm}{u_k^2}}
    \qty(\Dtmat{N}{\pm}{-u_k} \tmat{N}{\pm}{u_k} + \tmat{N}{\pm}{-u_k} \Dtmat{N}{\pm}{u_k})
    \label{eq:commutator-f}
    \,,
\end{align}
where in the first equality, we used Eq.~\eqref{eq:fermion-conjugate} and the symmetry of the normalization factor~\eqref{eq:normalization-factor-symm}, in the second equality, we used Eq.~\eqref{eq:tau-fermion-relation}, in the third and fifth equality, we used the relation proved by substituting $u= \pm u_k$ in Eq.~\eqref{eq:tau-fermion-relation}:
\begin{align}
    \tmat{N}{\pm}{u_k} f_k^{\dag} = f_k^{\dag} \tmat{N}{\pm}{-u_k} = 0
    \quad (k \in \modesetpositive{N}{\pm})
    \,,
\end{align}
and in the sixth equality, we use the anticommutation relation between the complex fermion $f_k$ and the Majorana fermion $\gamma_j$:
\begin{align}
    \label{eq:f-edge-anticomm}
    \qty{f_k^{\dag}, \gamma_j} & = \sqrt{\frac{2}{N}} e^{-\imi  (j-1) k}
    \quad(k \in \modeset{N}{\pm})
    \,.
\end{align}

Next, we will express the derivative of the logarithm of the transfer matrix in terms of the complex fermion~\eqref{eq:fermion-mode}.
We first consider the series expansion:
\begin{align}
    -\dv{u} \ln \tmat{N}{\pm}{u}
     & =
    - \frac{1}{\poly{N}{\pm}{u^2}} \tmat{N}{\pm}{-u} \Dtmat{N}{\pm}{u}
    =
    \sum_{r=1}^{\infty}
    \mathcal{H}_{\pm}^{(r)} u^{r-1}
    \,.
\end{align}
The charges $\mathcal{H}_{\pm}^{(r)}$ are given by the contour integral:
\begin{align}
    \mathcal{H}_{\pm}^{(r)}
    = &
    - \frac{1}{2\pi \imi} \oint \dd{u} \frac{1}{u^{r}} \frac{1}{ \poly{N}{\pm}{u^2}} \tmat{N}{\pm}{-u} \Dtmat{N}{\pm}{u}
    \nonumber \\
    = &
    - \frac{1}{2\pi \imi} \oint \dd{u} \frac{1}{u^{r}} \frac{1}{ \prod_{k \in \modesetpositive{N}{\pm}} (1 - u^2 / u_k^2) } \tmat{N}{\pm}{-u} \Dtmat{N}{\pm}{u}
    \nonumber \\
    = &
    - \frac{1}{2\pi \imi}
    \oint \dd{\varepsilon}
    \frac{\varepsilon^{2\abs{\modesetpositive{N}{\pm}} + r - 2 }}{\prod_{k \in \modesetpositive{N}{\pm}}
        (\varepsilon^2 - \varepsilon_k^2) }
    \tmat{N}{\pm}{-1/\varepsilon} \Dtmat{N}{\pm}{1/\varepsilon}
    \nonumber \\
    = &
    -\frac{1}{2} \sum_{k \in \modesetpositive{N}{\pm}} \frac{\varepsilon_k^{2\abs{\modesetpositive{N}{\pm}} + r - 3 }}{\prod_{l \in \modesetpositive{N}{\pm}, l \neq k} (\varepsilon_k^2 - \varepsilon_l^2)}
    ( \tmat{N}{\pm}{-1/\varepsilon_k} \Dtmat{N}{\pm}{1/\varepsilon_k}
    -(-1)^{r } \tmat{N}{\pm}{1/\varepsilon_k} \Dtmat{N}{\pm}{-1/\varepsilon_k} )
    \nonumber \\
    = &
    -\frac{1}{2} \sum_{k \in \modesetpositive{N}{\pm}} \frac{u_k^{-2\abs{\modesetpositive{N}{\pm}} - r + 3}}{\prod_{l \in \modesetpositive{N}{\pm}, l \neq k} (u_k^{-2} - u_l^{-2})}
    ( \tmat{N}{\pm}{-u_k} \Dtmat{N}{\pm}{u_k}
    -(-1)^{r } \tmat{N}{\pm}{u_k} \Dtmat{N}{\pm}{-u_k} )
    \nonumber \\
    = &
    \sum_{k \in \modesetpositive{N}{\pm}} \frac{u_k^{-r}}{\Dpoly{N}{\pm}{u_k^2}}
    \qty( \tmat{N}{\pm}{-u_k} \Dtmat{N}{\pm}{u_k}
    -(-1)^{r } \tmat{N}{\pm}{u_k} \Dtmat{N}{\pm}{-u_k} )
    \,,
\end{align}
where we use the variable transformation $\varepsilon = 1 / u$, $\varepsilon_k = 1 / u_k$, the integrals are taken around a clockwise closed path around $u=0$ and $\varepsilon=0$.
The radius of the closed path for the integration of $u$ is chosen sufficiently small to exclude all poles at $\varepsilon = \pm \varepsilon_k$, while the closed path for the integration of $\varepsilon$ is sufficiently large to encircle all poles at $\varepsilon = \pm \varepsilon_k$.
Note also that since the maximum power of $u$ in $\tmat{N}{\pm}{u}$ is $\floor{N/2}$, the integrand has no pole at $\varepsilon = 0$, with poles occurring only at $\pm \varepsilon_{k}$.
Also, we have used the relation:
\begin{align}
    \Dpoly{N}{\pm}{u_k^2}
    = - 2 u_k^{-1} \prod_{l \in \modesetpositive{N}{\pm}, l \neq k} (1 - u_k^2 / u_l^2)
    \qquad (k \in \modesetpositive{N}{\pm})
    \,.
\end{align}

Considering the derivative of the relation $\tmat{N}{\pm}{u} \tmat{N}{\pm}{-u} = u^{\delta_{\pm}} \poly{N}{\pm}{u^2}$, we have
\begin{align}
    \Dtmat{N}{\pm}{u_k} \tmat{N}{\pm}{-u_k} - \tmat{N}{\pm}{u_k} \Dtmat{N}{\pm}{-u_k}
    =
    u_k^{\delta_{\pm}} \Dpoly{N}{\pm}{u_k^2}
    \qquad (k \in \modesetpositive{N}{\pm})
    \,.
\end{align}
Then for $\mathcal{H}_{\pm}^{(2r)}$, we have
\begin{align}
    \mathcal{H}_{\pm}^{(2r)}
     & =
    \sum_{k \in \modesetpositive{N}{\pm}} \frac{u_k^{-2r}}{\Dpoly{N}{\pm}{u_k^2}}
    \qty( \tmat{N}{\pm}{-u_k} \Dtmat{N}{\pm}{u_k} -  \tmat{N}{\pm}{u_k} \Dtmat{N}{\pm}{-u_k} )
    =
    \sum_{k \in \modesetpositive{N}{\pm}} \varepsilon_k^{2r}
    \label{eq:even-ln-charge}
    \,.
\end{align}
For $\mathcal{H}_{\pm}^{(2r + 1)}$, using Eq.~\eqref{eq:commutator-f} we have
\begin{align}
    \mathcal{H}_{\pm}^{(2r + 1)}
    =
    \sum_{k \in \modesetpositive{N}{\pm}} \varepsilon_k^{2r + 1} s_k
    =
    \sum_{k \in \modesetpositive{N}{\pm}} (\varepsilon_k s_k)^{2r + 1}
    \,.
    \label{eq:odd-ln-charge}
\end{align}
Together with Eqs.~\eqref{eq:even-ln-charge} and~\eqref{eq:odd-ln-charge}, we have
\begin{align}
    \mathcal{H}_{\pm}^{(r)}
     & = \sum_{k \in \modesetpositive{N}{\pm}} \qty(\varepsilon_k s_k)^{r}
    \,.
\end{align}
Then we have
\begin{align}
    -\dv{u} \ln \tmat{N}{\pm}{u}
     & =
    \sum_{r=1}^{\infty} \mathcal{H}_{\pm}^{(r)} u^{r-1}
    =
    \sum_{k \in \modesetpositive{N}{\pm}} \frac{\varepsilon_k s_k}{1 - u \varepsilon_k s_k}
    =
    -\dv{u} \ln (\sqrt{N}\chi_0)^{\delta_{\pm}} \prod_{k \in \modesetpositive{N}{\pm}} \qty(1 - u \varepsilon_k s_k)
    \,,
\end{align}
where $\delta_+ = 0$ and $\delta_- = 1$.

Thus, we have the following differential equation:
\begin{align}
    \dv{u} \qty[ \ln \tmat{N}{\pm}{u} - \ln (\sqrt{N}\chi_0)^{\delta_{\pm}} \prod_{k \in \modesetpositive{N}{\pm}} \qty(1 - u \epsilon_k \qty(n_k - 1/2))]
    = 0\,,
\end{align}
where the initial condition is given by $\tmat{N}{\pm}{0} = (\sqrt{N}\chi_0)^{\delta_{\pm}}$ and $\epsilon_k = 2 \varepsilon_k$.
Thus, from the uniqueness of the ordinary differential equation, we have the formula for the transfer matrix~\eqref{eq:tau-fermion-prod}.

\section{Eigenstates of the SUSY Hamiltonians}
\label{app:eigenstate-susy-hamiltonian}
In this appendix, we derive the eigenstates of the SUSY Hamiltonians constructed from the squares of supercharges for even $N$.

From Eq~\eqref{eq:hamiltonian-supercharge-in-fermion-odd}, we have
\begin{align}
    \label{eq:susy-hamiltonian}
   \qty(\syk{N}{2p+1})^2
   =
    N \qty[
        \sum_{\substack{{\cal K} \subseteq \modesetpositive{N}{-} \\ |{\cal K}| = p}} \prod_{k \in {\cal K}} \epsilon_{k} \left( n_{k} - \frac{1}{2} \right)
    ]^2
    \,.
\end{align}
The corresponding eigenvalues are
\begin{align}
    E = N \qty[
        \sum_{\substack{{\cal K} \subseteq \modesetpositive{N}{-} \\ |{\cal K}| = p}} \prod_{k \in {\cal K}}
        \cot(k/2) s_k
    ]^2
    \,,
\end{align}
where $s_k \in \{+1,-1\}$.

In contrast to the supercharge itself, the SUSY Hamiltonian~\eqref{eq:susy-hamiltonian} preserves the fermion number parity, allowing eigenstates to be constructed within each parity sector.
The twofold degenerate vacua satisfy $(-1)^F \ket{0}_{\pm} = \pm \ket{0}_{\pm}$ where $F$ is the fermion number operator.
All eigenstates are then obtained by applying creation operators to these vacua: $\prod_{q=1}^{n}f_{k_q}^\dagger \ket{0}_{\pm}$ for $k_q \in \modesetpositive{N}{-}$ and $n < N/2$.

The number of single-particle modes $\{\epsilon_k\}_{k \in \modesetpositive{N}{-}}$ for the Majorana bilinear SUSY Hamiltonian $\chi_0 \syk{N}{3}$ is $\abs{\modesetpositive{N}{-}} = \frac{N}{2} - 1$.
However, the dimension of the Hilbert space is $2^{N/2}$.
This discrepancy in degrees of freedom is resolved by the twofold degenerate vacua~\cite{ozaki-katsura-prr-2025}.

\twocolumngrid

\bibliography{ref.bib}

\end{document}